\documentclass[11pt,a4paper]{article}
\pdfoutput=1
\usepackage{jheppub}
\usepackage{amsmath}
\usepackage{amssymb}
\usepackage{amsthm}
\usepackage{MnSymbol}
\usepackage{nicefrac}
\usepackage{amsfonts}
\usepackage{dsfont}
\usepackage[markup=nocolor]{changes}
\usepackage{graphicx}
\usepackage{bm}
\usepackage{xcolor}
\usepackage{slashed}
\usepackage{empheq}
\usepackage{hyperref}

\def \pt{\partial}

\def \STr{\textmd{STr}}
\def\grav{\textmd{grav}}

\begin{document}
\author[a]{Gustavo P. de Brito,} 
\author[a]{Astrid Eichhorn} 
\affiliation[a]{CP3-Origins, University of Southern Denmark, Campusvej 55, DK-5230 Odense M, Denmark}

\emailAdd{gustavo@sdu.dk}
\emailAdd{eichhorn@cp3.sdu.dk}

\title{Nonvanishing gravitational contribution to matter beta functions for vanishing dimensionful regulators}

\abstract{
We explore the effect of quantum gravity on matter within a Renormalization Group framework. 
First, our results provide an explicit example of how misleading conclusions can be drawn by analyzing the gravitational contributions to beta functions, instead of analyzing universal quantities, such as critical exponents, that can be extracted from the beta functions. This could be key to explain differences between perturbative studies and Functional Renormalization Group studies.
Second, we strengthen the evidence that asymptotically safe gravity could generate a predictive ultraviolet completion for matter theories with gauge interactions, even in the limit of vanishing dimensionful regulator function.
We also find that the situation can be more subtle with higher-order, gravity-induced matter interactions.
}

\maketitle

\section{Introduction}
The existence of a gravitational contribution to the running of gauge couplings has been the subject of intense scrutiny \cite{Robinson:2005fj,Pietrykowski:2006xy,Toms:2007sk,Ebert:2007gf,Toms:2008dq,Tang:2008ah,Daum:2009dn,Toms:2009vd,Toms:2010vy,Anber:2010uj,Ellis:2010rw,Toms:2011zza,Harst:2011zx,Folkerts:2011jz,Felipe:2011rs,Narain:2012te,Christiansen:2017gtg,Eichhorn:2017lry,Christiansen:2017cxa,Eichhorn:2017muy,Eichhorn:2019yzm,DeBrito:2019gdd,Bevilaqua:2021uzk}. On the one hand, such a contribution is non-universal already in the one-loop approximation and it has been claimed that the momentum-dependence of the gauge coupling does not receive a gravitational contribution \cite{Toms:2007sk,Ebert:2007gf,Anber:2010uj,Ellis:2010rw,Toms:2011zza,Felipe:2011rs,Narain:2012te}, unless the cosmological constant is nonvanishing \cite{Toms:2008dq,Toms:2009vd}; with other works claiming a nonzero contribution \cite{Tang:2008ah,Toms:2010vy,Bevilaqua:2021uzk}. 
On the other hand, from the perspective of asymptotically safe quantum gravity a gravitational contribution to the running gauge coupling may enable, e.g., the UV completion of the Abelian hypercharge sector of the Standard-Model (SM) \cite{Harst:2011zx,Christiansen:2017gtg,Eichhorn:2017lry}. Moreover, the gravitational contribution entails an enhancement in predictive power, resulting in a calculable value of the gauge coupling at low energies when starting from an asymptotically safe fixed point.
Within this setting, various calculations based on functional renormalization group (FRG) methods indicate a non-vanishing gravitational contribution to the scale dependence of the gauge couplings \cite{Daum:2009dn,Folkerts:2011jz,Harst:2011zx,Christiansen:2017gtg,Eichhorn:2017lry,Christiansen:2017cxa,Eichhorn:2019yzm,DeBrito:2019gdd}. Nevertheless, it is natural to ask whether such results are artifacts of a particular type of regularization scheme. 

One source of the apparent confusion between the perturbative line of research  \cite{Robinson:2005fj,Pietrykowski:2006xy,Toms:2007sk,Ebert:2007gf,Toms:2008dq,Tang:2008ah,Toms:2009vd,Toms:2010vy,Anber:2010uj,Ellis:2010rw,Toms:2011zza,Felipe:2011rs,Narain:2012te,Bevilaqua:2021uzk} and the asymptotically safe line of research \cite{Daum:2009dn,Harst:2011zx,Folkerts:2011jz,Christiansen:2017gtg,Eichhorn:2017lry,Christiansen:2017cxa,Eichhorn:2019yzm,DeBrito:2019gdd} lies in the different semantics in the perturbative and the asymptotically safe setting, see also the discussion in \cite{Bonanno:2020bil}:\\
In the perturbative setting, one refers to the physical momentum-dependence of couplings as their running. For logarithmic momentum dependence\footnote{A logarithmic momentum dependence is not the only possibility, even in perturbation theory. For instance, in gauge-Yukawa systems which are asymptotically safe strictly within perturbation theory \cite{Litim:2014uca}, couplings exhibit power-law running away from the asymptotically safe fixed point.}, this is mirrored in a corresponding RG scale dependence. In turn, logarithmic scale dependence corresponds to a universal one-loop contribution.\\
In the asymptotically safe setting, the RG-running refers to the dependence on the FRG scale $k$, which indicates, down to which momentum scale quantum fluctuations have been integrated out. This notion of running agrees with the first notion at one loop for the case where  only dimensionless couplings are involved.
Accordingly, it does not agree with the first notion in the context of gravity-matter systems. The physical momentum-dependence of couplings or rather vertex functions can also be evaluated with the FRG, see \cite{Denz:2016qks,Eichhorn:2018nda,Knorr:2019atm,Pawlowski:2020qer,Bonanno:2021squ,Knorr:2021niv,Fehre:2021eob} for examples of the momentum dependence of the graviton propagator, gravity-matter couplings and the gravitational interaction. Besides the physical momentum dependence, the RG dependence is of interest, because it contains two crucial pieces of information: first, it encodes whether a continuum limit can be taken (in the sense of sending a regularizing length scale to zero); second, it encodes whether the existence of a continuum limit imposes constraints on the values of couplings (i.e., whether predictions can be made from asymptotic safety).

A second source of confusion lies in the focus on the gravitational contribution on its own in \cite{Robinson:2005fj,Pietrykowski:2006xy,Ebert:2007gf,Toms:2010vy,Ellis:2010rw}, instead of the focus on universal quantities. It is well-known that beta functions are not universal quantities: even for canonically marginal couplings, non-universality sets in at three loops in perturbation theory. Therefore, to infer physical information from a beta function, a more careful analysis is necessary.
For instance, the existence of a gravity-induced interacting fixed point with a corresponding nontrivial critical exponent should be a universal statement. In order to make it, not just the gravitational contribution to the gauge coupling, but also the gravitational beta functions themselves need to be evaluated. Then, a change in the non-universal gravitational contribution to the gauge beta function may conceivably be compensated by a corresponding change in the gravitational beta functions, such that the critical exponent remains unaffected.\\
  
 Even if we restrict our attention to the FRG framework, the calculation of beta-functions in interacting gravity-matter systems involves several sources of non-universality. For instance, different choices of regulator, gauge-parameters and background can lead to different results for beta-functions. Gauge- and background-dependence can, at least in principle, be controlled by solving the flow equation along with modified Slavnov-Taylor and split-Ward identities.  
  
In this paper we focus on the regulator dependence of FRG calculations in gravity-matter systems. Our approach is based on a new type of (pseudo-)regulator, put forward in \cite{Baldazzi:2020vxk,Baldazzi:2021ijd}, which is characterized by an external interpolating parameter $a$. In the limit $a \rightarrow 0$, the (dimensionful) regulator can be removed and only universal contributions survive in the calculation of beta functions for canonically marginal couplings. This feature provides a highly nontrivial testing ground for the hypothesis that universal statements can be extracted by focusing on the critical exponents of the marginal couplings instead of individual contributions to a beta function.\\
 We caution that since the limit $a \rightarrow 0$ is a special case, in which the regulator function is removed, there is no requirement that physical information has to be independent of the value of $a$. Conversely, if physical information is found to be $a$-independent, we interpret this as a highly nontrivial indication of stability.\\

This paper is organized as follows: In Sec.~\ref{sec:nonunivres} we review how the effect of asymptotically safe quantum gravity on matter is evaluated and which intermediate steps of a calculation are affected by non-universality. In Sec.~\ref{sec:GravMat} we present results on gravity-matter systems with the novel regulator. In Sec.~\ref{sec::Results_Universal} we explore how universal results can arise and also discuss which aspects of gravity-matter systems do not exhibit universality in the vanishing-regulator limit. In Sec.~\ref{sec::WGB}, we explore how the vanishing regulator limit affects the fixed-point structure of induced interactions in the matter sector, with a particular focus on the \textit{weak gravity bound}. Finally, we conclude in Sec.~\ref{sec:conclusions}.  We present additional technical aspects in an appendix and we provide an ancillary notebook containing explicit expressions for the results used in this paper.

\section{Non-universal results from FRG calculations in gravity-matter systems}\label{sec:nonunivres}

Gravity contributes to the flow of marginal couplings in the matter sector in a non-universal way. In the FRG, part of this non-universality is due to the regularization procedure. This section explores this type of non-universality with a concrete example.

The FRG realizes the Wilsonian paradigm of renormalization by including an infrared (IR) regulator in the Boltzmann factor of the Euclidean path integral. This regulator depends on an IR cutoff scale $k$, the RG-scale, and suppresses modes with momenta smaller than $k$.\\
The central object in the FRG formalism is the flowing action $\Gamma_k$. This scale-dependent object is a modified Legendre transform of the coarse-grained generating functional. The functional $\Gamma_k$ interpolates between the full effective action $\Gamma$ when $k=0$ and the bare action $S_\textmd{bare}$ when $k \to \Lambda_{\textmd{UV}}$ (with $\Lambda_{\textmd{UV}}$ being a UV cutoff). 

The flowing action satisfies a formally exact flow equation, known as the Wetterich equation \cite{Wetterich:1992yh,Morris:1993qb}, which is given by
\begin{align}\label{eq::Flow_Equation}
	k \pt_k \Gamma_k = 
	\frac{1}{2} \STr\left[ \left( \Gamma_k^{(2)} + \textbf{R}_k \right)^{-1} 
	k \pt_k \textbf{R}_k\right] \,,
\end{align}
where $\STr$ denotes the super-trace, which traces over internal and spacetime indices, with an additional negative sign for Grassmann-valued fields, $\textbf{R}_k$ denotes the FRG regulator function, and  $\Gamma_k^{(2)}$ denotes the 2-point function derived from the flowing action. The flow equation \eqref{eq::Flow_Equation} can be used, at least within approximations (truncations), to derive beta-functions that define the flow of couplings w.r.t. to the RG-scale $k$.

In cases where the bare action is known, the FRG framework enables us to evaluate the full effective action by integrating the Wetterich equation from an appropriate initial condition. In cases where the bare action is not known, such as in the search for asymptotic safety, the FRG framework allows us to find candidates for fixed points by solving\footnote{ More precisely, the fixed-point equation is a requirement on the dimensionless counterpart of $\Gamma_k$, see App.~A of \cite{Eichhorn:2018nda} for a detailed discussion.} $k\,\partial_k\Gamma_k=0$, see \cite{Berges:2000ew,Pawlowski:2005xe,Gies:2006wv,Delamotte:2007pf,Rosten:2010vm,Dupuis:2020fhh} for reviews.

We write the regulator function $\textbf{R}_k$ in momentum space as a function of the four-momentum squared, $q^2$
\begin{align}
	\textbf{R}_k(q^2) =  q^2 \,r(q^2/k^2) \,\mathcal{Z}_k \,,
\end{align}
where $r$ is called the shape function and $\mathcal{Z}_k$ is a (generalized) wave-function renormalization factor. For non-scalar fields, $\mathcal{Z}_k$ also carries spacetime and internal indices; for fermionic fields, $\textbf{R}_k$ typically depends on $\slashed{q}$.
In position-space calculations, one has to replace the argument $q^2$ by an appropriate differential operator (e.g., $- \nabla^2$).\\
The choice of the shape function is only constrained by three requirements to ensure the appropriate mode suppression, in addition to a normalization $[q^2 r(q^2/k^2)]_{q^2=0} = k^2$. These are that i) high-momentum modes (with $q^2/k^2>1$) should be unsuppressed, ii) low-momentum modes (with $q^2/k^2<1$) should be suppressed, iii) the regulator should vanish for $k \rightarrow 0$ and it should diverge for $k^2 \rightarrow \infty$. Thus, $r(q^2/k^2) \rightarrow 0$ for $q^2/k^2>1$, $r(q^2/k^2)>0$ for $q^2/k^2<1$.\\
Different choices for the shape function $r$ can lead to different beta functions. This is a manifestation of the non-universality of beta functions: because they are not observables, they may depend on such unphysical choices.\\
There is, however, a specific class of contributions that are universal with respect to the choice of $r$. To identify these universal contributions, we consider threshold integrals which occur in the evaluation of the trace in the flow equation \eqref{eq::Flow_Equation}
\begin{align}\label{eq::thresh_Int}
	I_{\alpha,\beta}[r] = \!\int_0^{\infty}\!\! \frac{dy\,}{(4\pi)^2}  
	\frac{y^{\alpha}  \, r^\prime(y)}{ \left( 1 + r(y) \right)^{\beta} } \, ,
\end{align}
where $y = q^2/k^2$ is a dimensionless variable\footnote{Because the threshold integrals can be expressed purely in terms of this dimensionless combination, the beta functions of the dimensionless counterparts of couplings are autonomous, i.e., do not explicitly depend on $k$.}.
The threshold integrals arise, when the flow equation Eq.~\eqref{eq::Flow_Equation} is projected onto a particular field monomial to extract the corresponding beta function. This projection gives rise to one-loop diagrams with the appropriate external legs. The momentum integration in such a diagram (the trace in Eq.~\eqref{eq::Flow_Equation}) gives rise to one of the threshold integrals in Eq.~\eqref{eq::thresh_Int}.
In general, the result for $I_{\alpha,\beta} $ depends on the choice of $r$. However, for $\alpha = 0$, the threshold integrals are independent of the shape function,
resulting in \cite{Berges:2000ew,Codello:2008vh}
\footnote{This result can be inferred from Eq.~(A18) in Ref.~\cite{Codello:2008vh}. We use the correspondence $I_{\alpha\beta}[r] = -\frac{k^{-2(n+1-\beta)}}{32\pi^2} \Gamma[n] Q_n\left[ \frac{\pt_t R_k}{P_k^\beta}\right]$  (with $n = \alpha + \beta - 1$) to translate our threshold integrals into the $Q$-functionals used in Ref.~\cite{Codello:2008vh}.}
\begin{align}
	I_{0,\beta}[r] = - \frac{1}{(4 \pi)^2} \frac{1}{\beta-1} \, .
\end{align} 

This universal contribution does not appear in the gravitational contribution to the flow of an Abelian gauge coupling $g$ at leading order in the Newton coupling \cite{Daum:2009dn,Harst:2011zx,Folkerts:2011jz,Christiansen:2017gtg,Eichhorn:2017lry,Christiansen:2017cxa,Eichhorn:2019yzm,DeBrito:2019gdd}, which reads (see Sec.~\ref{sec::Setup} for details on the technical setup)
\begin{align}\label{eq::beta_g_grav}
	k \partial_k g |_{\grav} = -\frac{80\pi}{3} \Big(  I_{1,2}[r] -2\, I_{1,3}[r] \Big) \, G \, g \,.
\end{align}
$G = k^2 G_\textmd{N}$ represents the dimensionless Newton coupling, with $G_\textmd{N}$ denoting its dimensionful counterpart. 
Eq.~\eqref{eq::beta_g_grav} shows that the (1-loop) gravitational contribution to the flow of an Abelian gauge coupling is composed only of non-universal threshold integrals. The non-universality of Eq.~\eqref{eq::beta_g_grav} can be confirmed by explicit calculations with different choices of shape function. For example, using the Litim \cite{Litim:2001up} and exponential shape functions,
\begin{align}\label{eq::shape_examples}
	r^\textmd{Litim}(y) = \left( \frac{1}{y} - 1\right) \theta\left(1-y\right) \,
	\qquad \textmd{and} \qquad
	r^\textmd{exp}(y) = \frac{1}{e^{y} - 1} \,,
\end{align}
we get
\begin{align}\label{eq:betagG}
	k \partial_k g |_{\grav}^{\textmd{Litim}} = - \frac{5}{18\pi}  G \, g \,
	\qquad \textmd{and} \qquad
	k \partial_k g |_{\grav}^{\textmd{exp}} = - \frac{5}{6\pi}  G \, g \,.
\end{align}
We argue that although the results in Eq.~\eqref{eq:betagG} differ, the sign of the result is universal: The sign is determined by the difference $I_{1,2}[r] -2\, I_{1,3}[r]$ in Eq.~\eqref{eq::beta_g_grav}, which can be written as
\begin{eqnarray}
	I_{1,2}[r] -2\, I_{1,3}[r] = \int_0^{\infty} \frac{dy}{(4 \pi)^2} \frac{y\, r'(y)}{(1+r(y))^2}\left(1- \frac{2}{1+r(y)} \right).
\end{eqnarray}
To determine the sign of the integrand, we note that i) $r'(y)<0$ , ii) the integrand is peaked at $y \approx 1$ and iii) $r(1) <1$.  Condition i) follows, because the regulator must suppress IR modes and vanish for UV modes. Condition ii) follows because of the interplay of the mass-like IR suppression in the denominator and the UV suppression by $r'(y)$ in the numerator. Condition iii) follows because the normalization condition $[q^2 r(q^2/k^2)]_{q^2=0} = k^2$ implies $y r(y) \rightarrow  1$ for $y \rightarrow 0$, and because $r(y)$ is a monotonically decreasing function of $y$, thus $r(y)<y$ and $r(1)<1$.\\
Together, the conditions imply that the main contribution to the integral comes from $y \approx 1$, where $r'(y)<0$ and $1-2/(1+r(y))<0$. Thus, the overall sign of the integrand is positive. The sign of $k\, \partial_k\, g|_{\rm grav}$ is therefore negative.
\\
The same result was obtained in \cite{Folkerts:2011jz,Christiansen:2017cxa}, based on a kinematical identity.\\
We will show how further universal information is encoded in Eq.~\eqref{eq::beta_g_grav} in Sec.~\ref{sec:GravMat} below, using the ``vanishing regulators" introduced in \cite{Baldazzi:2020vxk,Baldazzi:2021ijd}.

The negative sign in Eq.~\eqref{eq:betagG} implies that the gravitational contribution that is $\sim G$ renders gauge couplings asymptotically free, thus solving the Landau pole/triviality problem in U(1) gauge theory and preserving the fundamental nature of non-Abelian gauge theories. In addition, if a screening contribution from matter is present in the beta function, the competition between the gravitational and the matter term induces an asymptotically safe fixed point with enhanced predictive power \cite{Harst:2011zx,Eichhorn:2017lry}.\\

We obtained the result in \eqref{eq::beta_g_grav} in a setup with vanishing cosmological constant. For non-vanishing cosmological constant $\bar{\Lambda}$, there are further non-universal contributions, as well as a universal one.
We expand $k \partial_k g |_{\grav}$ in powers of $\bar{\Lambda}$ to obtain the universal contribution which is proportional to the dimensionless product $G_\textmd{N}  \bar{\Lambda}$.
In this way, we find a vanishing result,
\begin{align}
	k \partial_k g |_{\grav}^{\textmd{univ}} = -\frac{320 \pi}{3} \Big( 2\,I_{0,3}[r] - 3\, I_{0,4}[r] \Big)  G_\textmd{N}  \bar{\Lambda} \,g = 0 \,.
\end{align}
This vanishing result relies on a cancellation of terms coming from different diagrams contributing to the gauge field anomalous dimension.

\section{Gravity-matter systems with vanishing regulators}\label{sec:GravMat}
\subsection{FRG with vanishing regulators}\label{sec::Vanishing_Reg}
We investigate the impact of quantum gravity on the flow of matter couplings using the class of ``vanishing regulators''  \cite{Baldazzi:2020vxk,Baldazzi:2021ijd}. This class of regulators has a shape function $\hat{r}_{a}$ that depends on an additional parameter $a$. This parameter modulates the amplitude of $\hat{r}_{a}$.
The interpolating shape function $\hat{r}_{a}$ is
\begin{align}
	\hat{r}_{a}(y) = a \, r(y) \,,
\end{align}
where $r(y)$ is a shape function satisfying all properties required for an FRG regulator. 
The parameter $a$ interpolates between the ``standard" shape-functions for $a \rightarrow 1$ and vanishing shape functions for $a \rightarrow 0$.
For the shape-functions defined in \eqref{eq::shape_examples}, we have
\begin{align}\label{eq::interp_shape}
	\hspace*{-.3cm}
	\hat{r}_{a}^\textmd{Litim}(y) = a \left( \frac{1}{y} - 1\right) \theta\left( 1-y \right) 
	\qquad \textmd{and} \qquad
	\hat{r}^\textmd{exp}_a(y) = \frac{a}{e^{y} - 1} \,.
\end{align}

The interpolating shape function $\hat{r}_{a}$ is a viable regulator for $a > 0$. If we set $a = 0$ from the beginning, the shape function vanishes, and $\hat{r}_{a}$ no longer is a viable regulator. Thus, this particular choice is referred to as vanishing (pseudo)-regulator. However, the limit $a \to 0$, if taken after evaluating the 1-loop integrals in the flow equation, generates non-vanishing contributions. Thus, the order of the limit $a \rightarrow 0$ and the integration over quantum fluctuations matters.

Using the interpolating shape function $\hat{r}_{a}^\textmd{Litim}(y)$ to evaluate the threshold integral defined in \eqref{eq::thresh_Int}, we find
\begin{align}\label{eq::thresh_Int_VanReg}
	I_{\alpha,\beta}[\hat{r}_{a}^\textmd{Litim}] = - \frac{a^{1-\beta}}{(4\pi)^2 (\alpha + \beta - 1)}  \,\, {}_2F_1\left(\beta, \alpha + \beta - 1 , \alpha + \beta ,1-a^{-1}\right) \,,
\end{align}
where ${}_2F_1$ is a hypergeometric function. This expression reproduces the results obtained with the standard Litim regulator given by \eqref{eq::shape_examples} in the limit $a \rightarrow 1$.
In the vanishing regulator limit, $a \to 0$, the threshold integral $I_{\alpha,\beta}$ behaves as
\begin{align}
	\lim_{a\to0}\, I_{\alpha,\beta}[\hat{r}_{a}^\textmd{Litim}] =
	\begin{cases}
		\qquad 0 \,\,, \qquad &\alpha > 0 \,,\\
		-\frac{1}{(4\pi)^2} \frac{1}{\beta - 1} \,\, ,\qquad &\alpha = 0 \,.
	\end{cases}
\end{align} 
For $\alpha < 0$ and $a \rightarrow 0$, the threshold integral $I_{\alpha,\beta}$ diverges in the IR. This IR divergence can be treated either by introducing a mass parameter in the propagator or by introducing a second regularization parameter $\epsilon$ as discussed in \cite{Baldazzi:2020vxk}.  This type of divergence is not relevant for the analysis performed in this work.

In the case of the interpolating $\hat{r}_{a}^\textmd{exp}(y)$, we are not aware of an analytical formula for the threshold integrals $I_{\alpha,\beta}[\hat{r}_{a}^\textmd{exp}]$ with arbitrary $\alpha$ and $\beta$. For the particular choices of $\alpha$ and $\beta$ that are relevant for this work, $I_{\alpha,\beta}[\hat{r}_{a}^\textmd{exp}]$ can be computed analytically, but the resulting expressions are lengthy and we shall not report them here.

In the limit $a \rightarrow 0$, where the dimensionful regulator is removed, we expect the resulting beta functions to be universal. Indeed,
in all cases where we explicitly computed $I_{\alpha,\beta}[\hat{r}_{a}^\textmd{exp}]$, the vanishing regulator limit ($a\to 0$) agrees with results obtained using $\hat{r}_{a}^\textmd{Litim}(y)$.\\

In summary, the vanishing regulator limit allows us to the isolate universal contributions in FRG calculations. By varying the interpolating parameter $a$, we can continuously deform ``standard'' FRG beta-functions into beta-functions that involve only universal contributions with respect to the shape function.

\subsection{Setup for the evaluation of beta functions \label{sec::Setup}} 
We study the gravitational contribution to the flow of SM-like interactions. To extract beta-functions from the flow equation \eqref{eq::Flow_Equation} we employ the following truncation for $\Gamma_k$
\begin{align}\label{eq:truncation}
	\Gamma_k =  \Gamma_k^{\textmd{matter}}  + \Gamma_k^{\grav} \,.
\end{align}
The matter sector includes a real scalar $\phi$, a Dirac spinor $\psi$, and an Abelian gauge field $A_\mu$. Our choice of truncation for $\Gamma_k^\textmd{matter}$ is given by
\begin{align}\label{eq::Gamma_Matter}
	\Gamma_k^{\textmd{matter}} = \!\int_x \sqrt{g} \,\left(  
	\frac{Z_\phi}{2} g^{\mu\nu}\pt_\mu\phi \pt_\nu \phi + \frac{\lambda}{4}\phi^4 + 
	i Z_\psi \bar{\psi} \slashed{\nabla} \psi + \frac{Z_A}{4} g^{\mu\alpha} g^{\nu\beta} F_{\mu\nu} F_{\alpha\beta}  + i  y\,\phi\, \bar{\psi}\psi \right) .
\end{align}
This matter sector contains key building blocks of Standard-Model like gauge-Yukawa theories, namely a quartic scalar interaction, a Yukawa interaction and a gauge coupling (related to the gauge-field wave-function renormalization). Because gravitational interactions are ``blind" to internal symmetries, many of our conclusions carry over to more complicated matter sectors.\\
In the gravitational sector, we truncate the dynamics to an Einstein-Hilbert term plus a gauge-fixing contribution. The gauge-fixing term includes an auxiliary background metric $\bar{g}_{\mu\nu}$, and is covariant with respect to background-gauge transformations. The background method is used to set up the calculation, such that the metric $g_{\mu\nu}$ is split into the background metric and a fluctuation field $h_{\mu\nu}$. The flowing effective action is given by
\begin{align}
	\Gamma_k^{\grav} = 
	-\frac{1}{16 \pi G_\textmd{N}} \int_x\sqrt{g} \, R  \, + \,
	\frac{1}{2\alpha_{\textmd{gf}}} \int_x \sqrt{\bar{g}} \,\bar{g}^{\mu\nu} F_{\mu}[h;\bar{g}] F_{\nu}[h;\bar{g}]  + \Gamma_{k}^{\rm ghost}\,,
\end{align}
with $F_{\mu}[h;\bar{g}] = \left( \delta^\alpha_\mu \bar{g}^{\nu\beta} - \frac{1 + \beta_{\textmd{gf}}}{4} \delta_\mu^\nu \bar{g}^{\alpha\beta}\right) \bar{\nabla}_\nu h_{\alpha\beta}$, where $\alpha_{\textmd{gf}}$ and $\beta_{\textmd{gf}}$ denote gauge-fixing parameters. Throughout this paper we consider the Landau gauge where $\alpha_{\textmd{gf}} \to 0$. $\Gamma_{k}^{\rm ghost}$ is the corresponding Faddeev-Popov ghost derived from the gauge-fixing function $F_{\mu}[h;\bar{g}]$.\\

Our analysis covers two different settings in the gravitational sector:\\
i) ``\textit{Standard gravity}'': In this setting, all metric degrees of freedom are included in the path integral, and the symmetry group (before breaking by the regulator) is the full diffeomorphism group.
Among the various ways of parameterizing and gauge-fixing metric fluctuations in this setting, we choose the linear split $g_{\mu\nu} = \bar{g}_{\mu\nu} + \kappa \,h_{\mu\nu}$ (with $\kappa = \sqrt{32\pi G_\textmd{N}}$) and set the gauge parameter $\beta_{\textmd{gf}}$ to $\beta_{\textmd{gf}} = 0$. The setting with linear split and Landau gauge  $\alpha_{\rm gf} \to 0$ is commonly used when exploring the interplay between gravity and matter couplings within the framework of asymptotically safe quantum gravity \cite{Eichhorn:2011pc,Eichhorn:2012va,Eichhorn:2016vvy,Eichhorn:2017eht,Eichhorn:2017lry,Eichhorn:2017sok,Eichhorn:2017ylw,Eichhorn:2018nda,Eichhorn:2018whv,Eichhorn:2019yzm,DeBrito:2019rrh,Eichhorn:2020kca,Eichhorn:2020sbo,deBrito:2020dta,deBrito:2021hde,Eichhorn:2021qet}, alternative gauge choices and parameterizations are explored in \cite{Eichhorn:2016esv,Christiansen:2017gtg,Eichhorn:2017als,DeBrito:2019gdd,deBrito:2021pyi}.  The impact of matter on the flow of gravitational couplings was investigated in \cite{Dona:2013qba,Dona:2014pla,Meibohm:2015twa,Oda:2015sma,Dona:2015tnf,Biemans:2017zca,Christiansen:2017cxa,Hamada:2017rvn,Pawlowski:2018ixd,Eichhorn:2018ydy,Alkofer:2018fxj,Eichhorn:2018akn,Wetterich:2019zdo,Burger:2019upn,Daas:2021abx,Laporte:2021kyp}.
\\
ii) ``\textit{Unimodular gravity}'': In this setting, the conformal mode is non-dynamical, and the symmetry group (before breaking by the regulator) is the group of transverse diffeomorphisms \cite{vanderBij:1981ym,Buchmuller:1988yn,Buchmuller:1988wx,Unruh:1988in,Henneaux:1989zc,Ellis:2010uc,deLeonArdon:2017qzg,Percacci:2017fsy}.
The metric is decomposed with the exponential split $g_{\mu\nu} = \bar{g}_{\mu\alpha}[ e^{\kappa h} ]^\alpha_{\,\,\,\nu}$, because that allows to easily implement a non-dynamical conformal mode by setting the gauge parameter $\beta_{\textmd{gf}} \to -\infty$. This choice of $\beta_{\textmd{gf}}$ enforces the trace of $h_{\mu\nu}$ to be constant, i.e., removed from among the fluctuating fields in the path integral. This property, combined with the exponential split of $g_{\mu\nu}$, enforces the unimodularity condition\footnote{There are multiple ways of implementing the unimodularity condition in quantum gravity \cite{deLeonArdon:2017qzg,Alvarez:2015sba,Bufalo:2015wda}. In particular, most of the perturbative studies in unimodular gravity are based on a version where the metric is redefined in terms of a ``densitized metric'' \cite{Alvarez:2015sba,Alvarez:2016uog,Gonzalez-Martin:2017fwz,Gonzalez-Martin:2018dmy,Herrero-Valea:2020xaq}. It is not clear whether different versions of unimodular quantum gravity are equivalent or not. The version of unimodular gravity used in this work is also referred as ``unimodular gauge'' \cite{Ohta:2015fcu,Percacci:2015wwa,Dona:2015tnf,deBrito:2020rwu,deBrito:2021pmw}.} $\det g_{\mu\nu} = \omega$ (where $\omega$ is a fixed density). Evidence for asymptotic safety in unimodular gravity was found in \cite{Eichhorn:2013xr,Benedetti:2015zsw,Eichhorn:2015bna,deBrito:2020rwu,deBrito:2020xhy}. The interplay between gravity and matter in unimodular asymptotically safe quantum gravity was investigated in \cite{Eichhorn:2015bna,DeBrito:2019gdd,deBrito:2020xhy}.

Our truncation for $\Gamma_k$ does not include a cosmological constant term. 
In the \textit{standard gravity} setting, this allows us to make contact with perturbative quantum gravity about a flat background.
In the \textit{unimodular gravity} setting, the cosmological constant naturally decouples from the system due to the unimodularity condition.\\
 At the practical level, the approximation of vanishing cosmological constant avoids technical issues related to the vanishing regulator limit in the presence of massive modes \cite{Baldazzi:2021ijd}. 


\subsection{Gravitational contribution to the flow of matter couplings  \label{sec::Grav_Contrib_Matter}} 

Our goal is to extract the beta functions of gauge, Yukawa, quartic-scalar and gravitational couplings in order to analyze the critical exponents of tentative fixed points in the vanishing regulator limit.

We compute the gravitational contribution to the flow of the gauge coupling $g$ according to
\begin{align}
	k \partial_k g|_{\grav} = \frac{1}{2} \eta_A^{\grav}\, g   = - f_g \, g \,,
\end{align}
where $\frac{1}{2}\eta_A^{\grav} = - f_g$ is the gravitational contribution to the anomalous dimension of the Abelian gauge field $A_{\mu}$, see also \cite{Daum:2009dn,Folkerts:2011jz,Christiansen:2017gtg,Eichhorn:2017lry,Christiansen:2017cxa,Eichhorn:2019yzm,DeBrito:2019gdd}.

For the gravitational contribution to the flow of the quartic scalar coupling $\lambda$ and Yukawa coupling $y$, we parameterize the gravitational contribution to their flow according to
\begin{eqnarray}
	k \partial_k \lambda|_{\grav} &=& 2\eta_\phi^{\grav} \,\lambda + \mathcal{D}_\lambda \lambda = -f_\lambda \,\lambda  \,,\\
	k \partial_k y|_{\grav}& =& \left( \eta_\psi^{\grav}+ \frac{1}{2} \eta_\phi^{\grav} \right)\, y  + \mathcal{D}_y \,y = -f_y \,y \,.
\end{eqnarray}
 We use $\eta_\phi^{\grav} $ and $\eta_\psi^{\grav} $ to denote the gravitational contribution to the anomalous dimensions of the scalar and fermion fields. The remaining parts,  $\mathcal{D}_\lambda \lambda $ and $\mathcal{D}_y \,y$, correspond to the direct contributions coming from diagrams involving gravity-matter vertices extracted from the quartic-scalar and Yukawa sectors in \eqref{eq::Gamma_Matter}.

In the SM, the Abelian gauge, the Higgs quartic, and the Yukawa couplings are irrelevant at the free fixed point. This fact prohibits a simple perturbative UV completion of the SM,  although it is not a settled fact that the SM as a whole suffers from a triviality problem. Separately, it is known that the scalar sector suffers from a triviality problem \cite{Frohlich:1982tw}, as does the Abelian gauge sector \cite{Gell-Mann:1954yli}; see however \cite{Gies:2020xuh} for the suggestion of an asymptotically safe version of QED.\\
If we assume that the perturbative Landau poles in the Higgs quartic and Abelian gauge sector of the SM translate into a triviality problem of the SM, then
the situation may change under the impact of quantum gravity: If $f_y>0$ and $f_g>0$, Abelian gauge and Yukawa couplings are relevant at their free fixed point and may exhibit interacting fixed points. Starting from these interacting fixed points, the infrared values of these couplings are calculable from first principles \cite{Harst:2011zx,Eichhorn:2017lry,Eichhorn:2017eht,Eichhorn:2017ylw,Eichhorn:2018whv,Alkofer:2020vtb}.
At the same time, if $f_{\lambda}<0$, the Higgs quartic coupling remains irrelevant at the free fixed point, translating into a Higgs mass in the vicinity of the experimental value \cite{Shaposhnikov:2009pv,Pawlowski:2018ixd,Eichhorn:2020sbo}.

At the fixed point at which gravity is interacting, but $g_{\ast}=0$, $\lambda_{\ast}=0$ and $y_{\ast}=0$, the quantities $f_g$, $f_{\lambda}$ and $f_y$ correspond directly to the critical exponents and therefore correspond to universal quantities. We stress that this requires that they are evaluated at the fixed-point values for $G = G_\textmd{N} k^{2}$, i.e., $f_g \big|_{G = G_{\ast}}$ is universal, but $f_g \big|_{G \neq G_{\ast}}$ is not. Further, we stress that universality does not necessarily imply regulator-independence within a truncated setup, as we consider here. However, it does imply that the vanishing-regulator limit should be shape-function independent.

We get the following results in the ``standard gravity" setting\footnote{We use self-written Mathematica codes based on the packages \textit{xAct} \cite{Brizuela:2008ra,Martin-Garcia:2007bqa,Martin-Garcia:2008yei}, \textit{FormTracer} \cite{Cyrol:2016zqb} and \textit{DoFun} \cite{Huber:2019dkb} to derive the beta functions.}
\begin{align}
	f_g &=  \frac{80 \pi }{3} \,\Big(  I_{1,2}[\hat{r}_a] - 2\,I_{1,3}[\hat{r}_a] \Big) \, G \,,\\
	f_\lambda &= \frac{16 \pi}{3} \, \Big( 17 \,I_{1,2}[\hat{r}_a] +2 \,I_{1,3}[\hat{r}_a] \Big) \,G \,,\\
	f_y&= \,\frac{\pi}{6} \,\Big( 220 I_{1,2}[\hat{r}_a] + 88\,I_{1,3}[\hat{r}_a] - 135 I_{1,5/2}[\hat{r}_a]  \Big) \, G \,.
\end{align}
To explore the limit $a \rightarrow 0$ and discuss the dependence on the shape function, we use $f_g$ as our main example, for which
\begin{align}
	f_g|^\textmd{Litim} &= -\frac{10}{6\pi} \, \left( \frac{ 2\,a}{(1-a)^2} + \frac{ a\,(a+1)\,\log (a)}{(1-a)^3}\right) \, G \,,\label{eq:fgLitim}\\  
	f_g|^\textmd{exp} &= -\frac{10}{6\pi} \, \left( \frac{ a}{1-a} + \frac{ a\,\log (a)}{(1-a)^2}\right) \, G \,.
\end{align}
In the limit $a \rightarrow 0$ with $G$ held fixed, $f_g \rightarrow 0$. This result is similar to results from perturbative studies, where the gravitational contribution to the gauge beta function vanishes if one uses a regularization scheme that does not introduce a mass-scale, e.g., dimensional regularization. As we will show below, however, the result that $f_g \rightarrow 0$ is an artefact of treating $G$ as a fixed external parameter.\\
In the limit $a \rightarrow 0$, the two expressions for the different shape functions agree (and the result is in that sense universal), but one might argue that they do so in a trivial fashion. Away from $a = 0$, at fixed values of $G$, the results do not agree.

We denote the same quantities in the \textit{unimodular gravity} setting by introducing the subscript ``UG'':
\begin{align}
	f_g|_{\textmd{UG}} &=  16 \pi \, \Big( I_{1,2}[\hat{r}_a] - 2\,I_{1,3}[\hat{r}_a] \Big) \, G \,,\\
	f_\lambda|_{\textmd{UG}} &= 48\pi  \, \Big( I_{1,2}[\hat{r}_a] +2 \,I_{1,3}[\hat{r}_a] \Big) \,G \,,\\
	f_y|_{\textmd{UG}} &= \frac{3\pi}{2} \,\Big( 12 \,I_{1,2}[\hat{r}_a] + 24\,I_{1,3}[\hat{r}_a] - 7 \,I_{1,5/2}[\hat{r}_a]   \Big) \, G \,.
\end{align}

\subsection{Flow of the Newton coupling}

To extract the universal content in $f_g, f_y$ and $f_{\lambda}$, we require a flow equation for the dimensionless Newton coupling $G$.  
We can extract the flow of $G$ within our setup by using the background field approximation, which results in the following expression
\begin{align}\label{eq::Flow_G}
	k \partial_k G &= 2G + \frac{8 \pi }{3} \Big( 23 \,I_{1,1}[\hat{r}_a] + c \, I_{1,2}[\hat{r}_a ] \Big) \,G^2  \,,
\end{align}
where $c = + 11$ in the \textit{standard gravity} setting and $c = -6$ in the \textit{unimodular gravity} setting.

\section{Universal results from FRG calculations in gravity-matter systems? \label{sec::Results_Universal}}

\subsection{Non-trivial $a \rightarrow 0$ limit for universal quantities}
We now demonstrate that i) the limit $a \rightarrow 0$ is no longer trivial, when taken for the universal quantity $f_g\Big|_{G = G_{\ast}}$ (corresponding to the critical exponent at the asymptotically free fixed point for the gauge coupling) and ii) this nontrivial limit $a \rightarrow 0$ agrees for the two choices of shape functions. To show the first point, we focus on the Litim shape function; to show the second point, we compare to the exponential shape function in a second step.

\begin{figure}[!t]
	\begin{center}
		\hspace*{-.5cm}
		\includegraphics[height= 6.cm]{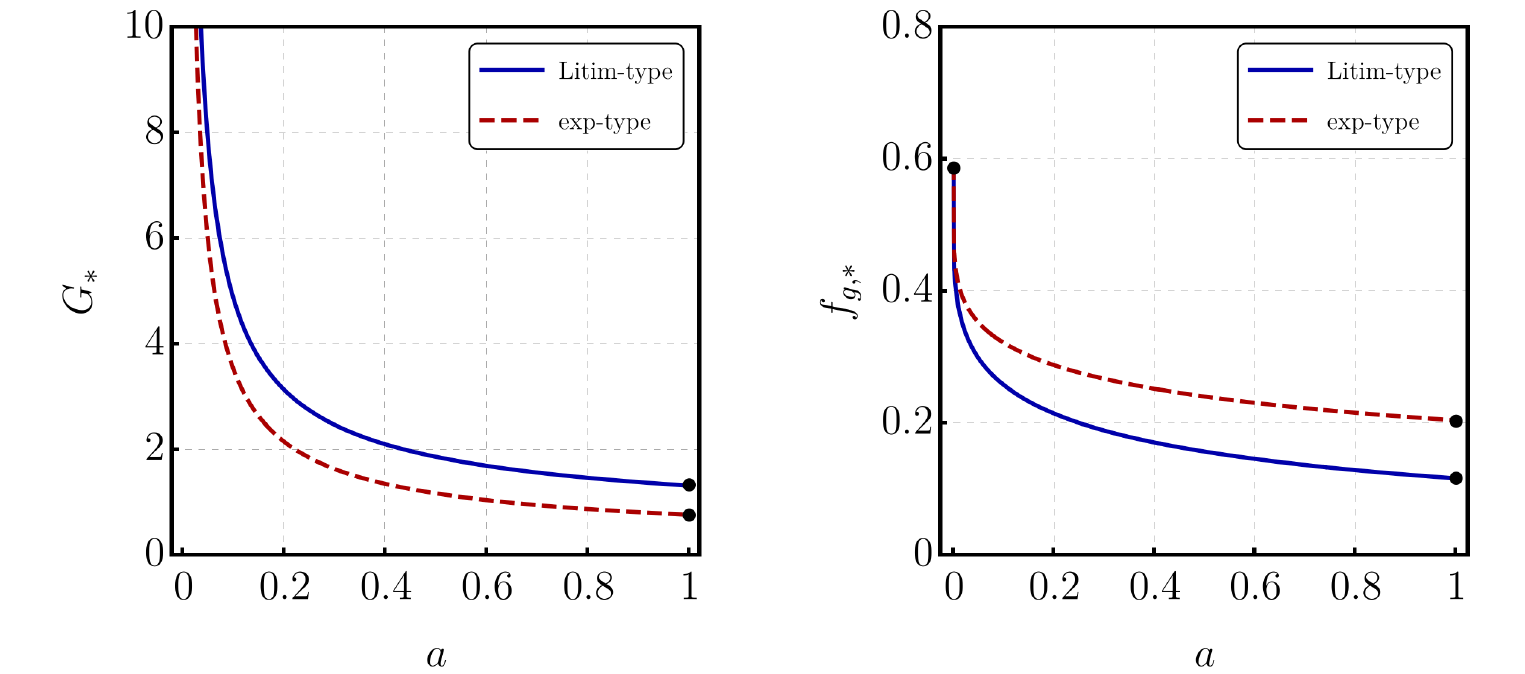} 
		\caption{Left panel: We show the fixed-point value for the Newton coupling, $G_{\ast}$, as a function of $a$ for the exponential (red, dashed line) and Litim-type (blue, continuous line) cutoff. Right panel: We show $f_{g, \ast} = f_{g}\Big|_{G = G_{\ast}}$, which is the critical exponent at the asymptotically free fixed point for the gauge coupling, as a function of $a$ for the two different shape functions. The result from both shape functions agrees for $a \rightarrow 0$ and is non-zero.}
		\label{fig::G_fixpt}
	\end{center}
\end{figure}

To show that $f_g\Big|_{G = G_{\ast}}$ is nontrivial in the limit $a \rightarrow 0$, it is key that the fixed-point value for $G$ depends on $a$. For the Litim shape function, the result is
\begin{align}
	G_*(a)^\textmd{Litim} = \frac{12 \pi \, (1-a)^2}{ (23 a-34) \,a\, \log (a) - 11\, (1-a)\, a } \,, \label{eq:G_FP_1,1}
\end{align}
which actually exhibits a divergence for $a \rightarrow 0$, cf.~left panel in Fig.~\ref{fig::G_fixpt},
\begin{align}
	G_*(a)= - \frac{6\pi}{17\,a \,\log(a)} \,+\,  \Omega(a)  \,, \label{eq:G_FP_small_a_Std}
\end{align}
where $\Omega(a)$ denotes a contribution that remains finite in the limit $a \rightarrow 0$. This scaling of $G_*(a)$ in the vanishing regulator limit was pointed out in \cite{Percacci:2020Talk}.
We can combine this with the expression for $f_g$ in Eq.~\eqref{eq:fgLitim} to obtain
\begin{align}
	f_{g}\Big|_{G = G_{\ast}}  \rightarrow \frac{10}{17}. \label{eq:fgLitimFP}
\end{align}
This non-zero and positive result shows that, if evaluated carefully, there is a gravitational contribution in the limit of vanishing regulator. The sign of the contribution renders the gauge coupling relevant at the free fixed point. At the same time, this opens up the possibility of an interacting fixed point, at which the gauge coupling would be irrelevant, which we come back to in Sec.~\ref{sec:NGFP}.\\

The result provides a cautionary example for those studies that evaluate the gravitational contribution to the flow of a matter coupling without accounting for a fixed point in the Newton coupling. For instance, the investigation in \cite{Folkerts:2011jz} did not investigate the behavior of $G$ for the choice of shape function that would result in $f_g=0$. Similarly, perturbative studies, e.g., in \cite{Robinson:2005fj,Pietrykowski:2006xy,Toms:2007sk,Ebert:2007gf,Toms:2008dq,Tang:2008ah,Toms:2009vd,Toms:2010vy,Anber:2010uj,Ellis:2010rw,Toms:2011zza,Felipe:2011rs,Narain:2012te,Bevilaqua:2021uzk}, treat $G$ as a fixed external parameter. We will come back to the specific case of the $\overline{\rm MS}$-limit from FRG calculations and the implications for previous studies in a separate work \cite{deBrito2022}.

\subsection{Universality check: Comparison of shape functions}
As a next step, we perform a universality check. We do not check a sufficient, but only a necessary condition, in confirming that the Litim shape function and the exponential shape function give the same result for $a \rightarrow 0$. For the exponential shape function, the fixed-point value for the Newton coupling is 
\begin{align}
	G_*(a)^\textmd{exp} = \frac{24 \pi  \,a\,(a-1)}{ \big( 46 (1- a) \log (1-a) - 23 (1 - a) \,\log (a) -22 a \big)\,a\log (a) - 46\, \Upsilon(a) } \,,  \label{eq:G_FP_2,1}
\end{align}
where
\begin{align}\label{eq::UpsilonFunction}
	\Upsilon(a) = a(1- a) \left( \textmd{Re}\Big( \textmd{Li}_2\left(1/a\right) - i  \pi \log(a) \Big) - \frac{\pi^2}{3}  \right)\,,
\end{align}
with $\textmd{Li}_2$ denoting a polylogarithm function.

The expression for $G_{\ast}$ diverges for $a \rightarrow 0$:
\begin{align}
	G_*(a)^\textmd{exp} = G_*(a)= - \frac{6\pi}{17\,a \,\log(a)} \,+\,  \Omega(a),
\end{align}
which is the same as for the Litim cutoff. Combining this with the expression for $f_g$ in the exponential parameterization yields
\begin{align}
	f_g\vert^{\rm exp} \rightarrow \frac{10}{17},
\end{align}
which agrees with the result \eqref{eq:fgLitimFP}.\\
We also observe that, despite the pole at $a=0$, the critical exponent associated with $G$ is finite for all $a\in[0,1]$. 
This follows because $\beta_G$ can be recast in the form $\beta_G = 2 \,G\left[ 1 - G/G_{\ast}(a)\right]$, which implies the critical exponent $\theta_G = 2$ for all values of $a$.

We also find that the quantitative variation of $f_g$ over the interval $a \in [0,1]$ is actually not more than a factor 3-6 (depending on the choice of shape function), with only minor changes until $a<0.1$, cf.~right panel of Fig.~\ref{fig::G_fixpt}.

\subsection{Unimodular gravity}
\begin{figure}[t]
	\begin{center}
		\hspace*{-.5cm}
		\includegraphics[height= 6.cm]{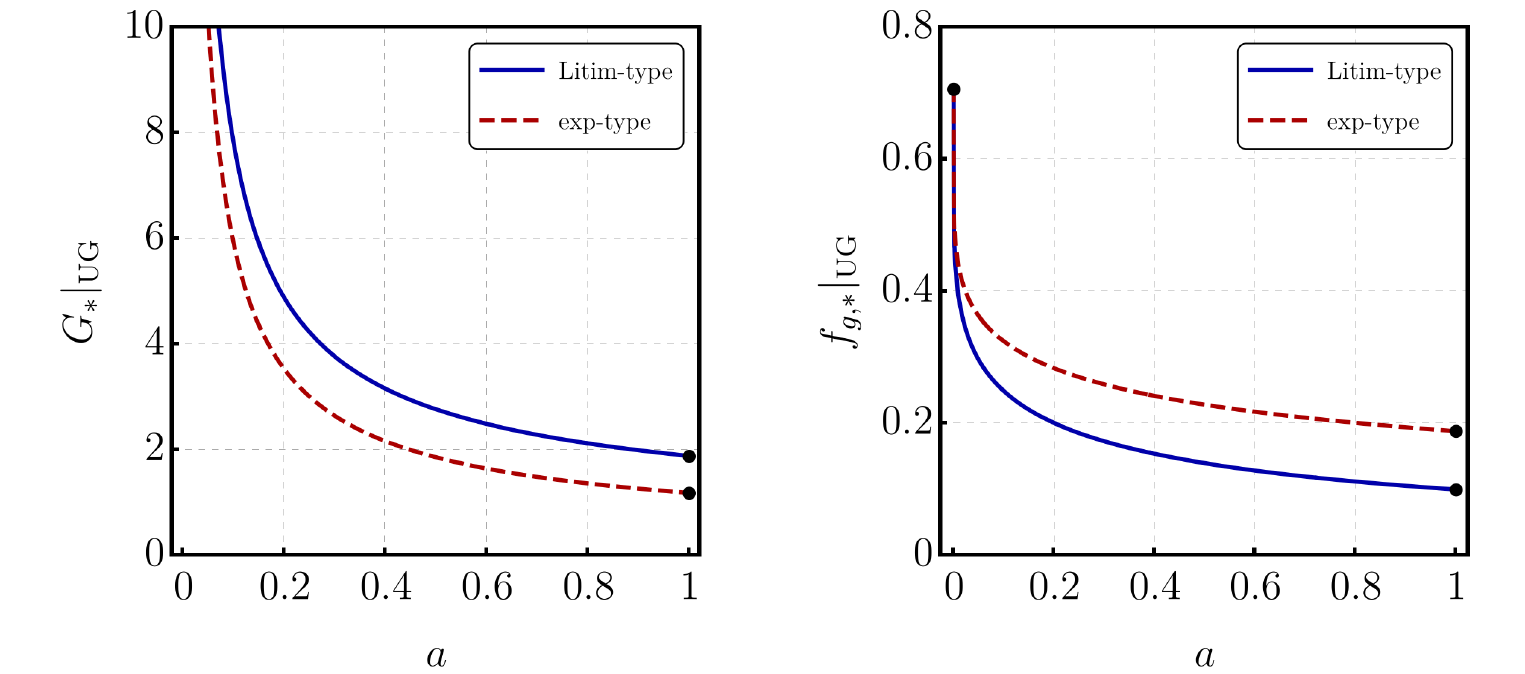} 
		\caption{Left panel: We show the fixed-point value for the Newton coupling, $G_{\ast}$, as a function of $a$ for the exponential (red, dashed line) and Litim-type (blue, continuous line) cutoff for unimodular gravity. Right panel: We show $f_{g, \ast} = f_{g}\Big|_{G = G_{\ast}}$, which is the critical exponent at the asymptotically free fixed point for the gauge coupling, as a function of $a$ for the two different shape functions. The result from both shape functions agrees for $a \rightarrow 0$ and is non-zero.}
		\label{fig::G_fixpt2}
	\end{center}
\end{figure}

We repeat the analysis done above for the unimodular setting. In this case,
using Eq.~\eqref{eq::interp_shape} to evaluate the threshold integral in Eq.~\eqref{eq::Flow_G}, we find
\begin{align}
	G_*(a)|_\textmd{UG}^\textmd{Litim} &= \frac{12 \pi \, (1-a)^2}{ (23 a-17) \,a\,\log (a) + 6\,(1-a)\,a } \,,  \label{eq:G_FP_1,2}
\end{align}
for the Litim-type shape-function, $\hat{r}_a^\textmd{Litim}$, and
\begin{align}
	G_*(a)|_\textmd{UG}^\textmd{exp} &= \frac{24 \pi  \,a\,(a-1)}{  \big(46 (1-a) \log (1-a) - 23 (1-a) \,\log (a) + 12 a \big) \,a\log (a) - 46\, \Upsilon(a) }  \label{eq:G_FP_2,2}
\end{align}
for the exponential shape-function $\hat{r}_a^\textmd{exp}$. The function $\Upsilon(a)$ is again defined in terms of a polylogarithm function, see Eq.~\eqref{eq::UpsilonFunction}.
We plot $G_*(a)$ in the left panel of Fig.~\ref{fig::G_fixpt2}.

In the vanishing regulator limit ($a\to0$), the fixed-point value $G_*$ diverges for both shape functions
\begin{align}
	G_*(a)|_\textmd{UG} &= - \frac{12\pi}{17\,a \,\log(a)} \,+\, \Omega_\textmd{UG}(a)  \,,  \label{eq:G_FP_small_a_UG} 
\end{align}
with $\Omega_\textmd{UG}(a)$ denoting a contribution that is finite in the limit $a\to0$. Despite the pole at $a=0$, the critical exponent is again finite and $\theta_G=2$ for all $a$.

The gravitational contribution to the flow of the Abelian gauge coupling with the interpolating shape functions in Eq.~\eqref{eq::shape_examples} reads
\begin{align}
	f_g|_{\textmd{UG}}^\textmd{Litim} &= - \frac{1}{\pi}  \, \left( \frac{ 2\,a}{(1-a)^2} + \frac{ a\,(a+1)\,\log (a)}{(1-a)^3}\right) \, G  \,,\\
	f_g|_{\textmd{UG}}^\textmd{exp} &= -\frac{1}{\pi} \, \left( \frac{ a}{1-a} + \frac{ a\,\log (a)}{(1-a)^2}\right) \, G \,.\
\end{align}
In Fig.~\ref{fig::G_fixpt2} we show how $f_{g}$ depends on $a$, when evaluated at the fixed-point value $G_{\ast}$. We observe the existence of non-vanishing and sign-preserving contributions for all values $a \in [0,1]$.

For small values of the interpolating parameter $a$, $f_g$ behaves according to
\begin{align}
	f_g|_{\textmd{UG}} &= \left(- \frac{1}{\pi}  \, a \,\log (a)    +  \Delta^{(g)}_\textmd{UG}(a) \right) \, G  \label{eq:f_g_small_a_UG} \,,
\end{align}
with $\Delta^{(g)}_\textmd{UG}(a)$ denoting a contribution that approaches zero faster than $a \log(a)$ in the limit $a \to 0$.

We can again take the vanishing regulator limit ($a \to 0$) in two different ways. The first possibility treats $G$ as a fixed parameter, yielding
\begin{align}
	f_g|_{\textmd{UG}}^{G_\textmd{fixed}}(a\to0) = 0 \,.
\end{align}
This way of taking the limit $a \to 0$ neglects the nontrivial $a$-dependence of $G$. 
The second possibility takes the $a$-dependence of $G$ into account. 
In particular, when evaluated at the gravitational fixed point where $G=G_*(a)$, the combination of Eq.~\eqref{eq:G_FP_small_a_UG} and Eq.~\eqref{eq:f_g_small_a_UG} leads to non-vanishing $f_g$ in the limit $a \to 0$ also for the unimodular setting:
\begin{align}
	f_{g,*}|_{\textmd{UG}}(a\to0) = \frac{12}{17} \,.
\end{align}
This result is again independent of the choice of Litim- vs.~exponential shape function. Again, the gauge coupling is relevant at the free fixed point under the impact of quantum gravity -- in this case, in its unimodular incarnation. This result highlights how important it is to focus on universal quantities (such as the critical exponent $f_{g,\, \ast}$), instead of calculating the gravitational contribution to a beta function with $G$ as a free parameter, see, e.g., \cite{Gonzalez-Martin:2017bvw}.

\subsection{Interacting fixed point in the gauge coupling}\label{sec:NGFP}

In the presence of charged matter, the flow of $g$ also receives a 1-loop contribution from a vacuum polarization diagram with a matter loop. 
We focus on the case of a charged Dirac fermion (thus adding the gauge connection to the Dirac operator in the flowing action Eq.~\eqref{eq::Gamma_Matter}) and
compute this contribution using the FRG, resulting in a universal contribution of $g^3/12\pi^2$. The resulting beta function for $g$ at one loop is then  given by
\begin{align}
	k \pt_k g =  - f_g \, g +\frac{1}{12\pi^2} g^3\,.
\end{align}
If $f_g > 0$, this beta function has an IR-attractive non-Gaussian fixed-point at $g_* = \sqrt{12\pi^2 \,f_{g,*}}$ with critical exponent $\theta_g = - f_{g,*}$ (and an IR-repulsive Gaussian fixed point with critical exponent $\theta_g =  f_{g,*}$ that we discussed above).
The fixed-point value itself is not a universal quantity, but the existence of a fixed point is a universal piece of information. In fact, $g_{\ast}$ remains finite and non-zero in the vanishing-regulator limit
\begin{align}
	g_*(a\to0) = 2\pi \sqrt{30/17} \approx 8.3
	\qquad \textmd{and} \qquad
	g_*|_{\textmd{UG}}(a\to0) = 12\pi \sqrt{1/17} \approx 9.1
\end{align}
The critical exponent at this fixed point, being just the negative of the critical exponent of the free fixed point, approaches a non-zero limit for $a \rightarrow 0$ in which the result from both shape functions agrees.

It is rather nontrivial that the limit $a\rightarrow 0$ exhibits such a high degree of stability, given that the regulator vanishes in this limit.
Therefore, there is no a priori need for a fixed point to persisting in this limit and its continued existence is remarkable.\\
We iterate that stability is exhibited by universal quantities, namely $f_{g}$ evaluated at $G_*$, which corresponds to critical exponents. 
In contrast, the gravitational contributions $f_g$ vanish when the limit $a \rightarrow 0$ is taken at fixed $G$, i.e., when we consider a non-universal quantity.

\subsection{Quartic scalar and Yukawa coupling}

\begin{figure}[t]
	\begin{center}
		\hspace*{-.45cm}
		\includegraphics[height= 6.cm]{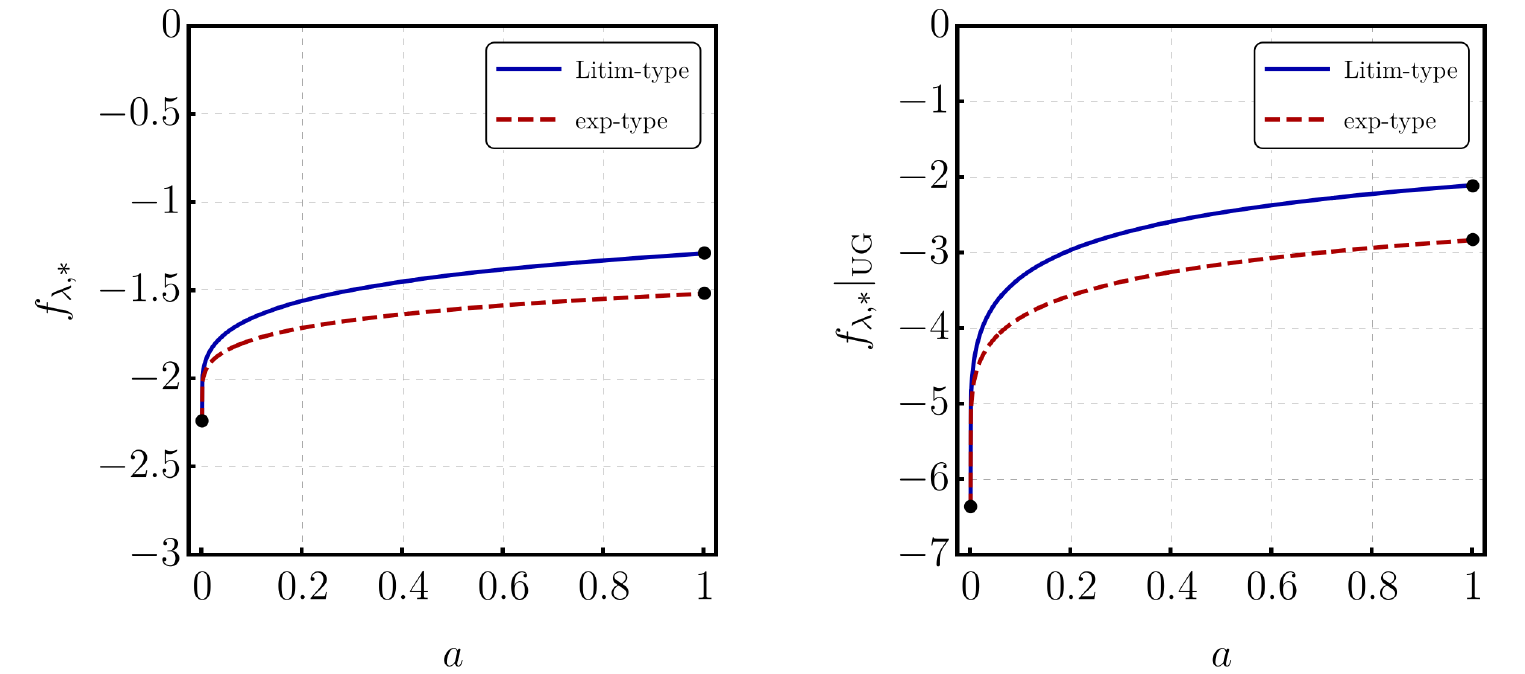} 
		\caption{Left panel: We show $f_{\lambda, \ast}$, corresponding to the critical exponent at the free fixed point in standard gravity for the exponential (red, dashed line) and Litim-time (blue continuous line) cutoff.  Right panel: We show $f_{\lambda, \ast}$, corresponding to the critical exponent at the free fixed point in unimodular gravity for the two cutoffs. }
		\label{fig::f_lambda_fixpt}
	\end{center}
\end{figure}

\begin{figure}[t]
	\begin{center}
		\hspace*{-.5cm}
		\includegraphics[height= 6.cm]{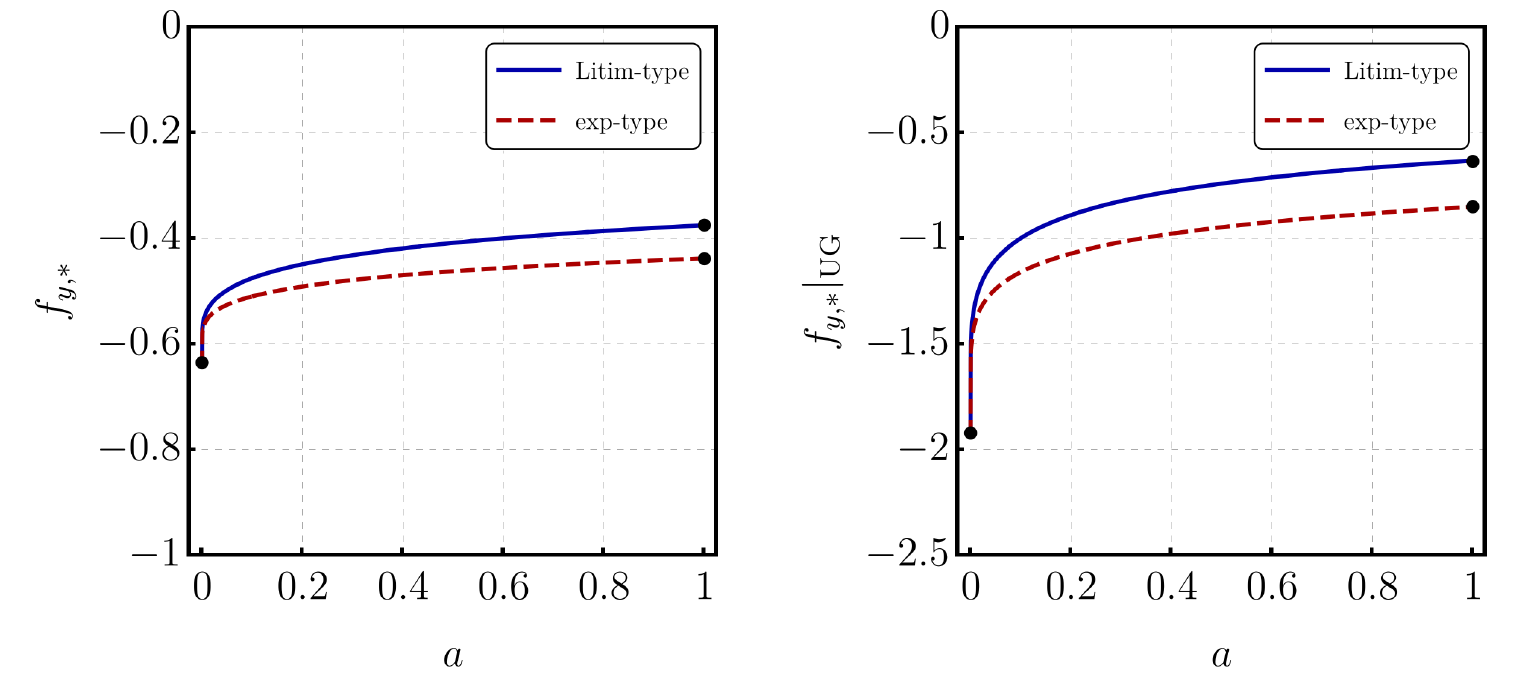} 
		\caption{Left panel: We show $f_{y, \ast}$, corresponding to the critical exponent at the free fixed point in standard gravity for the exponential (red, dashed line) and Litim-time (blue continuous line) cutoff.  Right panel: We show $f_{y, \ast}$, corresponding to the critical exponent at the free fixed point in unimodular gravity for the two cutoffs.}
		\label{fig::f_y_fixpt}
	\end{center}
\end{figure}

We perform a similar analysis for the quartic and Yukawa couplings, i.e., for $\lambda$ and $y$. 
For small values of the interpolating parameter $a$, the gravitational contributions $f_\lambda$ and $f_y$ behave as
\begin{align}
	f_\lambda&= \left( \frac{19}{3\pi} \, a \,\log (a)  +  \Delta^{(\lambda)}(a) \right) \, G  \label{eq:f_lambda_small_a_Std} \,,\\  
	f_\lambda|_{\textmd{UG}} &= \left( \frac{9}{\pi} \, a \,\log (a)    +  \Delta^{(\lambda)}_\textmd{UG}(a) \right) \, G  \label{eq:f_lambda_small_a_UG} \,,
\end{align}
\begin{align}
	f_y&= \left( \frac{173}{96\pi} \, a \,\log (a)  +  \Delta^{(y)}(a) \right) \, G  \label{eq:f_y_small_a_Std} \,,\\  
	f_y|_{\textmd{UG}} &= \left( \frac{87}{32\pi}  a \,\log (a)    +  \Delta^{(y)}_\textmd{UG}(a) \right) \, G  \label{eq:f_y_small_a_UG} \,,
\end{align}
with $\Delta^{(\lambda)}(a)$, $\Delta^{(\lambda)}_\textmd{UG}(a)$, $\Delta^{(y)}(a)$ and $\Delta^{(y)}_\textmd{UG}(a)$ approach zero faster than $a \log(a)$ in the limit $a \to 0$.
Taking the vanishing regulator limit with $G$ evaluated at $G_*(a)$, we get nonzero contributions in the limit $a\to0$,
\begin{align}
	f_{\lambda,*}(a\to0) = - \frac{38}{17} 
	\qquad \textmd{and} \qquad 
	f_{\lambda,*}|_{\textmd{UG}}(a\to0) = - \frac{108}{17} \,,
\end{align}
for the quartic-scalar coupling, and 
\begin{align}
	\hspace*{-.15cm}
	f_{y,*}(a\to0) = - \frac{173}{272}
	\qquad \textmd{and} \qquad 
	f_{y,*}|_{\textmd{UG}}(a\to0) = -\frac{261}{136} \, ,
\end{align}
for the Yukawa coupling. In Figs.~\ref{fig::f_lambda_fixpt} and \ref{fig::f_y_fixpt} we show $f_{\lambda,*}$ and $f_{y,*}$ within the range $a \in [0,1]$. Both $f_{\lambda,*}$ and $f_{y,*}$ are negative in this range. These signs are consistent with the results at $a=1$ and indeed no sign changes occur over the range $a \in [0,1]$. For the Yukawa coupling, it is known that the introduction of a negative cosmological constant, which is beyond the scope of our study, causes a change in sign in $f_y$ \cite{Eichhorn:2016esv,Eichhorn:2017eht}, which is at the heart of a set of tentative phenomenological consequences of asymptotic safety \cite{Eichhorn:2017ylw,Eichhorn:2018whv,Eichhorn:2020kca,Alkofer:2020vtb}.

In the Yukawa sector, we can also analyze the effective gravitational contribution $f_y^ \textmd{eff}$ which we obtain in the presence of a non-Gaussian fixed point for the Abelian gauge coupling $g$, cf.~\cite{Eichhorn:2018whv}. If $\psi$ couples to $A_\mu$, we can write the 1-loop flow equation for the Yukawa coupling according to
\begin{align}
	k \pt_k y =- f_y \,y +  \frac{5}{16\pi^2} y^3 -  \frac{3}{8\pi^2} g^2 y \,,
\end{align}  
with the coefficients $5/16\pi^2$ and $-3/8\pi^2$ being 1-loop universal in the usual sense, and being independent of the shape function in the FRG. At the gravity-induced non-Gaussian fixed point $g_* = \sqrt{12\pi^2 f_{g,\ast}}$, the flow of $y$ is determined by an effective scaling dimension $f^{\textmd{eff}}_y$
\begin{align}
	k \pt_k y|_{g=g^*} = - f^\textmd{eff}_y \,y+\frac{5}{16\pi^2} y^3   \,,
\end{align}  
with
\begin{align}
	f^\textmd{eff}_y = f_y + \frac{9}{2} f_g \,,
\end{align}
denoting the effective gravitational contribution to the flow of $y$. In this scenario, the condition for a gravity-induced UV-completion in the Yukawa sector is given by $f^\textmd{eff}_y > 0$, which can be achieved even if $f_y < 0$. 

\begin{figure}[t]
	\begin{center}
		\hspace*{-.5cm}
		\includegraphics[height= 6.cm]{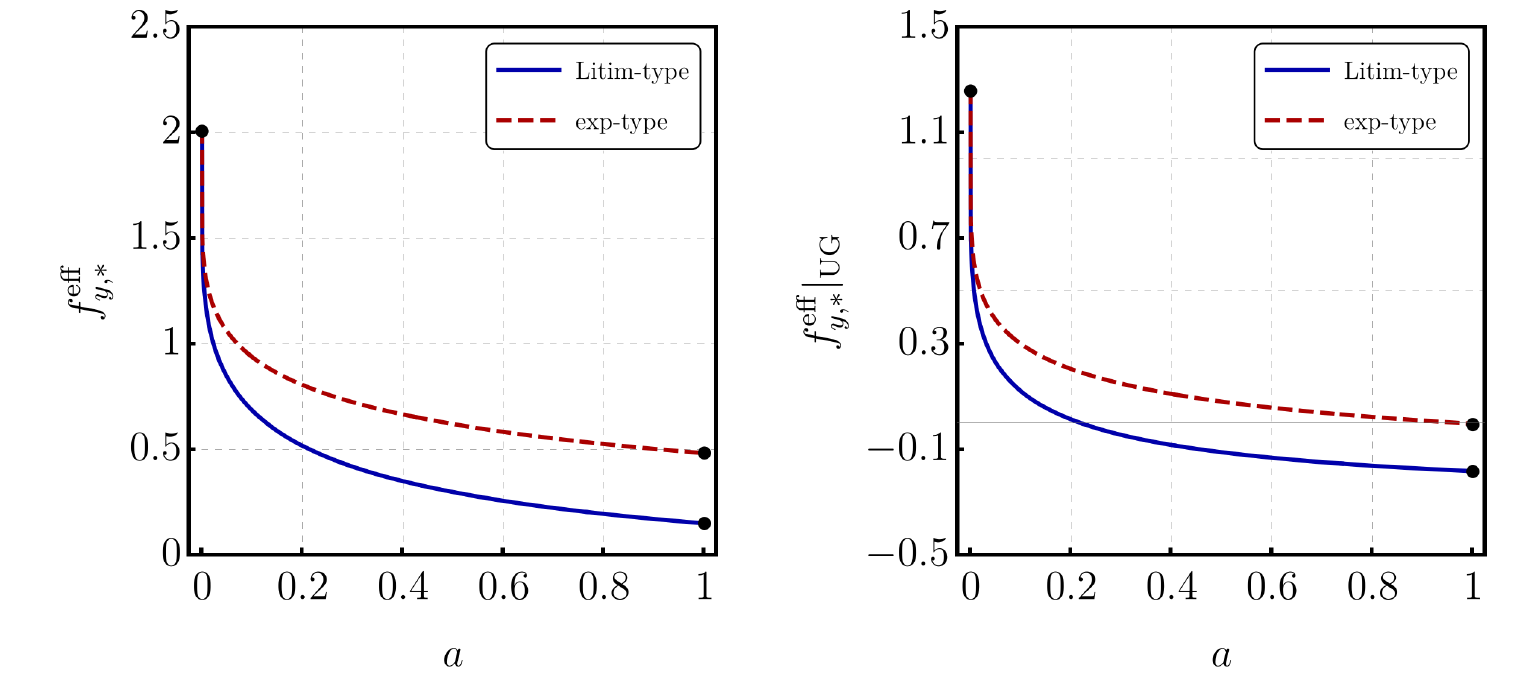} 
		\caption{Left panel: We show the effective scaling dimension at the fixed point with finite gauge and vanishing Yukawa coupling, $f_{y,\ast}^{\rm eff}$ in standard gravity for the exponential (red, dashed line) and Litim-type (blue continuous line) regulator. Right panel: We show $f_{y,\ast}^{\rm eff}$ in unimodular gravity for the two shape functions.}
		\label{fig::f_y_Eff_fixpt}
	\end{center}
\end{figure}

In Fig.~\ref{fig::f_y_Eff_fixpt}, we show $f_{y,*}^\textmd{eff}$ within the range $a \in [0,1]$. In the case of \textit{standard gravity}, $f_{y,*}^\textmd{eff}>0$ for all values of $a$ within this range. In the case of \textit{unimodular gravity}, however,  $f_{y,*}^\textmd{eff}$ does not have a definite sign.
We remind the reader that it is not a requirement that the $f$'s are $a$-independent in small truncations, as we consider here. These truncations neglect contributions to the $f$'s, coming, e.g., from higher-order gravitational couplings, as explored in \cite{DeBrito:2019gdd}, or from non-minimal couplings. In particular, there is no guarantee that a finite $a \rightarrow 0$ limit exists at all, given that it is a limit in which the regulator simply vanishes. Therefore, we consider the change of sign in $f_{y,\ast}^{\rm eff}\vert_{\rm UG}$ as a behavior that one might have expected. Instead, the existence of a nontrivial $a \rightarrow 0$ limit in $f_{y,\ast}^{\rm eff}\vert_{\rm UG}$, and the remarkable stability of all other critical exponents are in our view strong indications for truncations which are already more converged than one might have a priori expected.

In the limit $a \to 1$, corresponding to standard FRG shape functions, the results computed with \textit{standard gravity} and \textit{unimodular gravity} have the opposite sign
\begin{align}
	\hspace*{1.3cm}
	f^\textmd{eff}_y|_{\textmd{Std.}}^\textmd{Litim}(a\to1) &= \frac{13}{36 \pi } \, G \, 
	\hspace*{.31cm}\qquad \text{and} \qquad
	f^\textmd{eff}_y|_{\textmd{Std.}}^\textmd{exp}(a\to1)  =  \frac{157-90 \log (2)}{48 \pi } \,G\,,\\
	f^\textmd{eff}_y|_{\textmd{UG}}^\textmd{Litim}(a\to1) &=  -\frac{3}{10 \pi } \,G\,
	\hspace*{.1cm}\qquad \text{and} \qquad
	\hspace*{0cm}f^\textmd{eff}_y|_{\textmd{UG}}^\textmd{exp}(a\to1) = -\frac{42 \log (2)-29}{48 \pi } \,G \,.
\end{align}
In the vanishing regulator limit, with $G = G_*(a)$, the results computed with \textit{standard gravity} and \textit{unimodular gravity} agree on the sign of $f_{y,*}^\text{eff}$, namely
\begin{align}
	\hspace*{-.15cm}
	f^\textmd{eff}_{y,*}|_{\textmd{Std.}}(a\to0) =  \frac{547}{272}
	\qquad \textmd{and} \qquad 
	f^\textmd{eff}_{y,*}|_{\textmd{UG}}(a\to0) = \frac{171}{136} \, .
\end{align}

\subsection{Interpretation and outlook}
In light of the results presented in this section, we offer a new point of view for the disparity between the perturbative results on the presence of a gravitational contribution in the beta function \cite{Toms:2007sk,Ebert:2007gf,Anber:2010uj} and the functional RG results \cite{Daum:2009dn,Folkerts:2011jz,Christiansen:2017gtg,Eichhorn:2017lry,Christiansen:2017cxa,Eichhorn:2019yzm,DeBrito:2019gdd}. The former focus on a non-universal quantity\footnote{In a second step, perturbative studies explore the physical running of the gauge coupling as a function of the momentum. This is not our focus here; we focus on the results at the level of the beta functions.}, namely the gravitational contribution to the beta function of the gauge coupling at fixed $G$. The latter focuses on a universal quantity, namely the critical exponent associated with the gauge coupling. Here, we have shown within the FRG framework, how these two quantities behave quite differently, with the non-universal quantity vanishing and the universal quantity being non-zero. This highlights a potential source of disparity between the results: they evaluate different quantities, therefore a direct comparison is misleading.\\
Thus, it is conceivable that they can be reconciled and brought into a qualitative agreement if the same universal quantity is evaluated in the perturbative approach.
This may be achieved based on \cite{Baldazzi:2020vxk}, where an FRG- regulator has been proposed that mimics the $\overline{\textmd{MS}}$ scheme in a particular limit. Similar to the limit $a \rightarrow 0$, it is conceivable that in the $\overline{\textmd{MS}}$-limit, the gravitational fixed-point value diverges, while $f_{g,\lambda,y}$ vanish, such that the associated critical exponent stays finite and nonzero.

\section{Weak- gravity bound in the vanishing regulator limit} \label{sec::WGB}

Gravity-matter interactions can induce higher-order self-interactions in the matter sector \cite{Eichhorn:2011pc,Eichhorn:2012va,Eichhorn:2016esv,Christiansen:2017gtg,Laporte:2021kyp}.  
Consequently, there is no fixed point that is fully non-interacting in the matter sector, but interacting in the gravitational sector.
A specific class of gravity-induced matter self-interactions leads to a bound on the strength of the gravitational couplings at the fixed point, which is referred to as \textit{weak gravity bound} \cite{Eichhorn:2016esv,Eichhorn:2017eht,Christiansen:2017gtg,Eichhorn:2019yzm,deBrito:2021pyi,Eichhorn:2021qet}. It limits the value of the Newton coupling to lie below a critical value, beyond which induced matter self-interactions no longer feature a real-valued fixed point.
In this section, we investigate the \textit{weak gravity bound} in the vanishing- regulator limit.

We consider, as an example, gravity coupled to a scalar field, initially just through the kinetic term
\begin{align}
	\Gamma_k^\textmd{kin.} 
	= \frac{Z_\phi}{2} \int_x \sqrt{g} \, g^{\mu\nu} \pt_\mu \phi \pt_\nu \phi \,. 
\end{align}
The RG-flow generates interactions that are compatible with the global symmetries of the kinetic term, namely $\phi \mapsto - \phi$ ($\mathbb{Z}_2$-symmetry) and $\phi \mapsto \phi + \epsilon$ (shift-symmetry), where $\epsilon$ denotes a constant parameter. 
Therefore, in the minimal setup for gravity coupled to a scalar field, we can consistently set to zero all operators that violate $\mathbb{Z}_2$- and/or shift-symmetry (e.g., $\phi^n$ self-interactions\footnote{This is consistent with the previous section, in which the fixed point at which we investigated the gravitational contribution lies at $\lambda_{\ast}=0$.}), while the RG-flow generates interactions that are compatible with these symmetries (e.g., $g^{\mu\nu} g^{\alpha\beta} \pt_\mu \phi \pt_\nu \phi \pt_\alpha \phi \pt_\beta \phi$).

Here, we investigate the viability of a fixed-point regime within this minimal setup. For that, we extend our truncation  for $\Gamma_k$ in Eq.~\eqref{eq:truncation} by adding the induced interaction, see also \cite{deBrito:2021pyi}
\begin{align}
	\Gamma_k^\textmd{induced} = \frac{\bar{g}_\phi}{8}\int_x \sqrt{g} \, g^{\mu\nu} g^{\alpha\beta} \pt_\mu \phi \pt_\nu \phi \pt_\alpha \phi \pt_\beta\phi \,.
\end{align}

In this truncation, the flow of the dimensionless coupling $g_\phi = k^{4} Z_\phi^{-2} \, \bar{g}_\phi$ reads
\begin{align}\label{eq:flow_g_phi}
	k \pt_k g_\phi = 4 g_\phi \,+\, \beta_{g_\phi}^{(2,0)} \, g_\phi^2 
	\,+\, \beta_{g_\phi}^{(1,1)} \, g_\phi G \,+\,  \beta_{g_\phi}^{(0,2)} \, G^2 \,,
\end{align}
where 
\begin{align}
	&\beta_{g_\phi}^{(2,0)} = - 3 I_{2,2}[\hat{r}_a] - 5 I_{2,3}[\hat{r}_a]  \,, \\
	&\beta_{g_\phi}^{(1,1)} = \frac{32 \pi}{3} \big( 10 \, I_{1,2}[\hat{r}_a] - I_{1,3}[\hat{r}_a] - 3\,I_{1,4}[\hat{r}_a] \big) \,, \\
	&\beta_{g_\phi}^{(0,2)} = - \frac{512 \pi^2}{9} \big(40\,I_{0,3}[\hat{r}_a] + I_{0,5}[\hat{r}_a] \big) \,,
\end{align}
for the \textit{standard gravity} setting, and 
\begin{align}
	&\beta_{g_\phi}^{(2,0)}|_\textmd{UG} = - 3 I_{2,2}[\hat{r}_a] - 5 I_{2,3}[\hat{r}_a]  \,, \\
	&\beta_{g_\phi}^{(1,1)}|_\textmd{UG} = 16 \pi \big( I_{1,2}[\hat{r}_a] + 10 \, I_{1,3}[\hat{r}_a] - 18\,I_{1,4}[\hat{r}_a] \big) \,, \\
	&\beta_{g_\phi}^{(0,2)}|_\textmd{UG} = - 128 \pi^2 \big( 13\,I_{0,3}[\hat{r}_a] 
	- 30\,I_{0,4}[\hat{r}_a] +  36\,I_{0,5}[\hat{r}_a] \big) \,.
\end{align}
for the \textit{unimodular gravity} setting.

Since Eq.~\eqref{eq:flow_g_phi} is quadratic in $g_\phi$, it necessarily has two zeros in the complex plane, namely
\begin{align}
	g_{\phi,*}^{(\pm)} = \frac{-\left( 4 + G\, \beta_{g_\phi}^{(1,1)} \right) 
		\pm \sqrt{ \left(4 + G\, \beta_{g_\phi}^{(1,1)} \right)^2 - 4\, G^2 \beta_{g_\phi}^{(2,0)} \beta_{g_\phi}^{(0,2)} } }{2 \beta_{g_\phi}^{(2,0)} } \,.
\end{align}
For $G=0$, they both lie on the real line. In this case, the fixed point defined by the positive sign is a Gaussian fixed point, i.e., $g_{\phi,*}^{(+)}|_{G=0} = 0$. At $G \neq 0$, it is shifted away from zero;
thus, we refer to $g_{\phi,*}^{(+)}$ as a shifted Gaussian fixed point when $G \neq 0$.

For $G>0$, the existence of real fixed points depends on the sign of $\Delta = \left(4 + G\, \beta_{g_\phi}^{(1,1)} \right)^2 - 4\, G^2 \beta_{g_\phi}^{(2,0)} \beta_{g_\phi}^{(0,2)}$. A (non-degenerate) pair of real fixed points requires $\Delta >0$. 
Within truncated FRG flows, $\Delta >0$ only holds, if the gravitational fixed-point remains in the \textit{weak gravity} regime. More precisely, $G$ must be smaller than a critical values $G_\textmd{crit}$. This constraint on the fixed-point values of the gravitational couplings is the \textit{weak gravity bound}.

Using the Litim-type regulator to compute $G_\textmd{crit}$, we find
\begin{align}
	G_\textmd{crit.}(a\to1) \approx 1.4 \,,
\end{align}
in the \textit{standard gravity} setting, and
\begin{align}
	G_\textmd{crit.}|_\textmd{UG}(a\to1) \approx 2.9 \,,
\end{align}
in the \textit{unimodular gravity} setting.

In Ref.~\cite{deBrito:2021pyi}, we have seen that the fixed-point value $G_*$ for a gravity-scalar system (in the \textit{standard gravity} setting) lies above the critical value $G_{\text{crit.}}$, indicating that other types of matter fields are necessary to reconcile the fixed-point value $G_*$ with the \textit{weak gravity bound}. In particular, the inclusion of vector fields may reduce the value of $G_*$ such that it becomes smaller than $G_{\text{crit.}}$. Therefore, we add to our truncation a set of $N_A$ Abelian gauge fields
\begin{align}
	\Gamma_k^{\textmd{Abelian-gauge}} = \sum_{i=1}^{N_A} \frac{1}{4} \int_x g^{\mu\alpha} g^{\nu\beta} F_{\mu\nu}^i F_{\alpha\beta}^i \,.
\end{align}
We work in an approximation where we consider the impact of the Abelian gauge fields on the flow of the Newton coupling, but discard any form of back-reaction, i.e., we do not account for induced self-interactions of the gauge field, nor for induced gauge-scalar interactions. 
Within our truncation, the flow of the Newton coupling (obtained via the background field approximation) results in the following equation
\begin{align}
	k\pt_k G = 2 G + \frac{8\pi}{3} \big( (21+4N_A) I_{1,1}[\hat{r}_a] + c \,I_{1,2}[\hat{r}_a] \big)  G^2 \,.
\end{align}
with $c = + 11$ in the \textit{standard gravity} setting and $c = -6$ in the \textit{unimodular gravity} setting.

\begin{figure}[!t]
	\begin{center}
		\hspace*{-.5cm}
		\includegraphics[height= 6.cm]{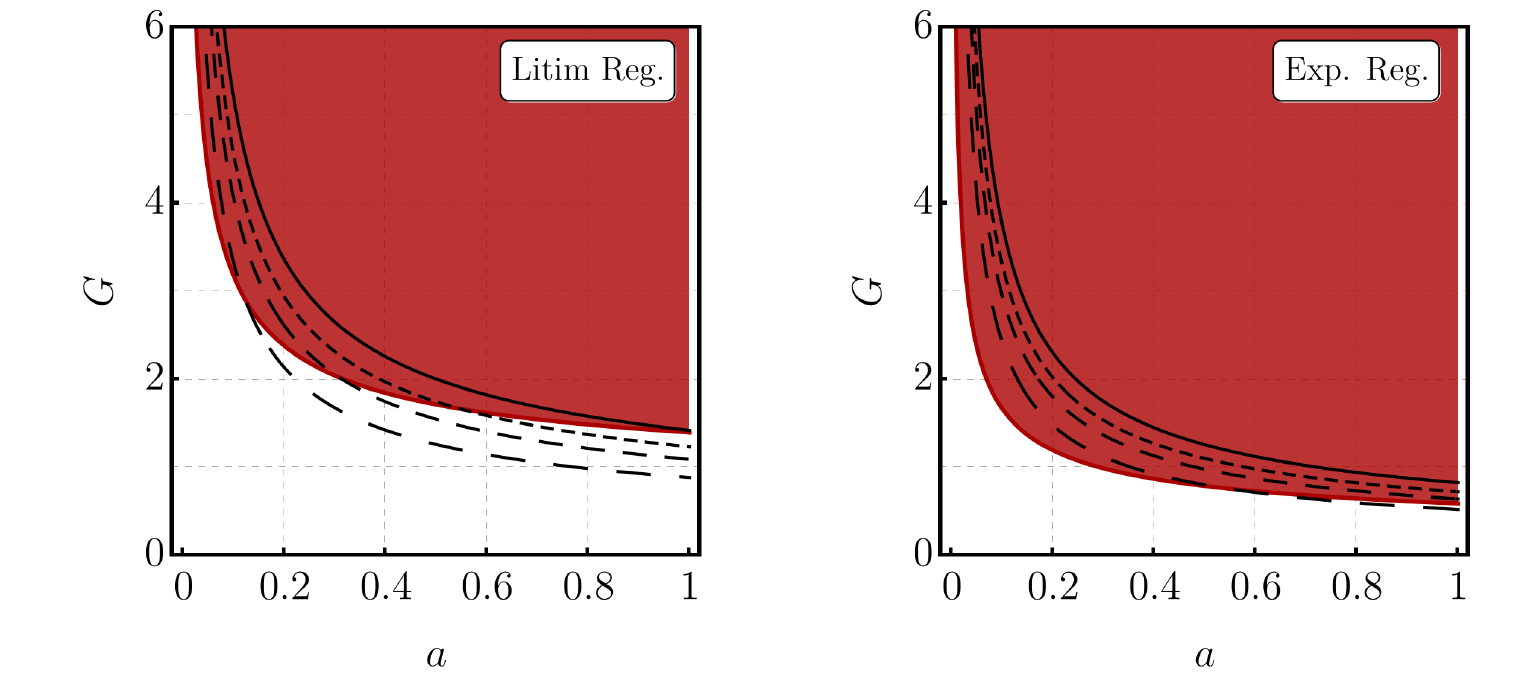} 
		\caption{\footnotesize {We show the \textit{weak gravity bound} in standard gravity for the Litim-type (left panel) and exponential-type (right panel) regulators. The red region is excluded by the \textit{weak gravity bound}. The black lines indicate the fixed-point values $G_*(a)$. Different lines correspond to different number ($N_A$) Abelian-gauge fields added to our truncation. From top to bottom: $N_A=0,1,2,4$.} }
		\label{fig::WGB_Standard}
	\end{center}
\end{figure}

\begin{figure}[!t]
	\begin{center}
		\hspace*{-.5cm}
		\includegraphics[height= 6.cm]{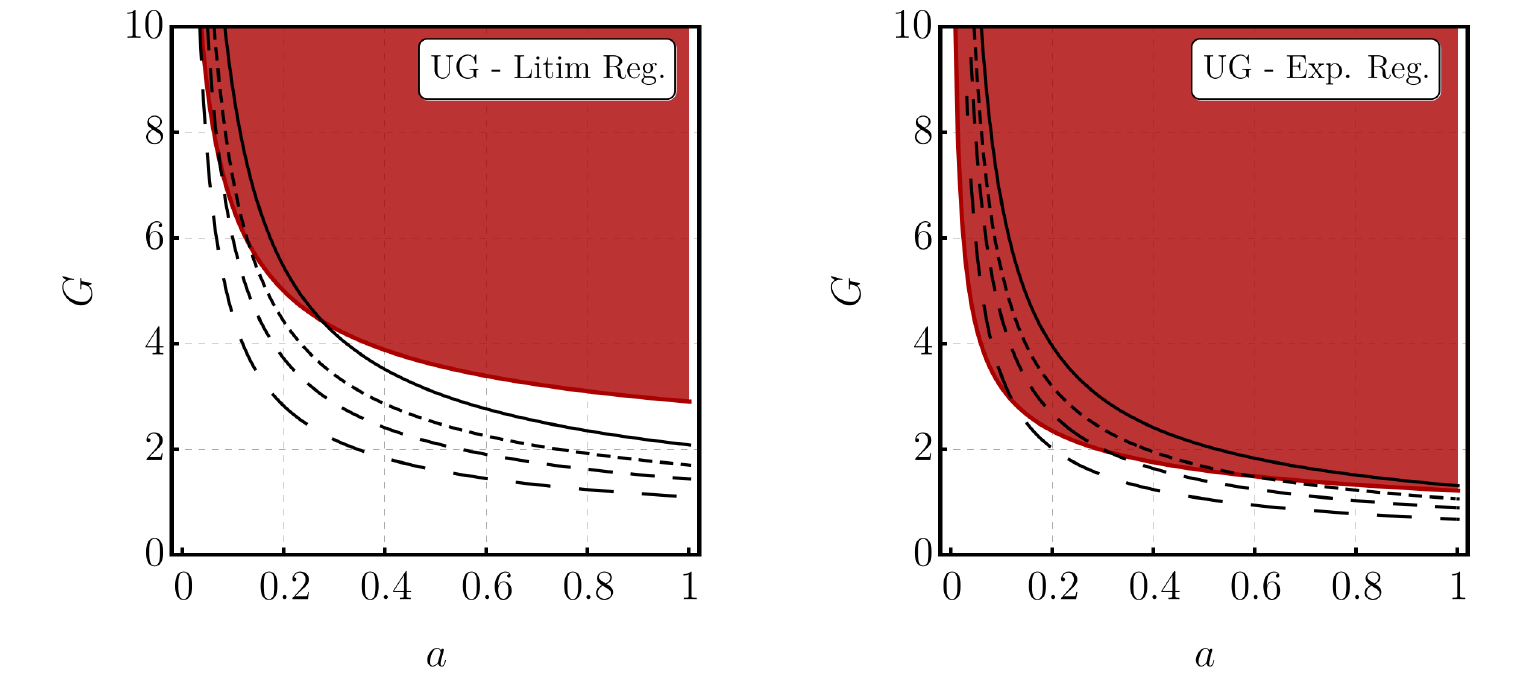} 
		\caption{\footnotesize {We show the \textit{weak gravity bound} in unimodular gravity for the Litim-type (left panel) and exponential-type (right panel) regulators. The red region is excluded by the \textit{weak gravity bound}. The black lines indicate the fixed-point values $G_*(a)$. Different lines correspond to different number ($N_A$) of Abelian-gauge fields added to our truncation. From top to bottom: $N_A=0,1,2,4$.} }
		\label{fig::WGB_Unimodular}
	\end{center}
\end{figure}

For the interpolating shape function $\hat{r}_a$ we naturally get an $a$-dependent result for $G_\textmd{crit.}$. As a result, the \textit{weak gravity bound} leads to an excluded region in the $a \times G$ plane. In Figs.~\eqref{fig::WGB_Standard} and \eqref{fig::WGB_Unimodular}, we show the regions that are excluded by the \textit{weak gravity bound} in the \textit{standard} and \textit{unimodular} settings, respectively.

The precise form of $G_\textmd{crit.}(a)$ depends on the choice of shape-function.
Such differences between shape functions are expected, because $G_{\rm crit}$ is not a physical quantity, similarly to fixed-point values. However, when we approach the vanishing regulator limit ($ a \to 0$), we observe the universal scaling
\begin{align}\label{eq:Gcrit_scaling}
	G_\textmd{crit.}(a) \sim \frac{1}{\sqrt{a}} \,.
\end{align}
This scaling is common between standard and unimodular gravity.
In particular, $G_\textmd{crit.}$ goes to infinity when $a \to 0$. This result suggests the naive interpretation that the \textit{weak gravity bound} would disappear in the vanishing regulator limit. 

However, we should compare the scaling of $G_\textmd{crit.}(a)$ with the scaling of the fixed-point value for $G$ in the vanishing regulator limit\footnote{In our setup, the induced interactions do not affect the beta function for $G$. Thus, a real fixed-point value for $G$ can exist, even if the induced interactions do not feature a real fixed point at this value of $G$.}. 
As discussed in Sec. \ref{sec::Results_Universal} (see also App.~\ref{app::Beyond1Loop}), the fixed-point value $G_*(a)$ behaves like
\begin{align}\label{eq:Gfp_scaling}
	G_*(a) \sim - \frac{1}{a \,\log(a)} \,, 
\end{align}
in the vanishing regulator limit. Therefore, it also goes to infinity in the limit $a \to 0$.

The existence of a real fixed point for the induced coupling $g_\phi$ in the vanishing regulator limit depends on whether $G_*(a)$ goes to infinity faster than $G_\textmd{crit.}(a)$. This is indeed the case, as we can see from the limit
\begin{align}
	\quad \lim_{a \to 0^+} \frac{G_*(a)}{G_\textmd{crit.}(a)} \to \infty \,.
\end{align}
Thus, we find no real fixed points for the induced coupling $g_\phi$ in the vanishing regulator limit.

Let us comment on the generality of these results
\begin{itemize}
	\item The results depend only on the scaling of $G_\textmd{crit.}(a)$ and $G_*(a)$ in the vanishing regulator limit, but not on numerical prefactors. Within our truncation, the scaling properties of $G_\textmd{crit.}(a)$ and $G_*(a)$ are not qualitatively affected by other choices. For example, we observe the same behavior for different settings in the gravitational sector and different choices of shape function.
	\item The qualitative picture remains the same, when we include the impact of additional matter fields on the flow of $G$. In Fig.~\eqref{fig::WGB_Standard} and \eqref{fig::WGB_Unimodular} we also show the behavior of the fixed point $G_*(a)$ once we include the impact of Abelian-gauge fields.  
	\item The results are not a peculiarity of scalar fields. For example, we explicitly checked that the induced coupling associated with the $F^4$-term for Abelian gauge fields has an associated \textit{weak gravity bound} with critical value $G_\textmd{crit.}$ scaling as \eqref{eq:Gcrit_scaling} in the vanishing regulator limit\footnote{This result was obtained within the \textit{standard gravity} setting. In the \textit{unimodular setting}, the analysis also requires the inclusion of $( F_{\mu\nu} \tilde{F}^{\mu\nu} )^2$ in our truncation. This is related to the $\beta_\textmd{gf}$-dependence of the \textit{weak gravity bound} in the Abelian gauge sector, see~\cite{Eichhorn:2021qet} for further discussion on this point.}.
\end{itemize}

There could be several interpretations of this result:\\ 
First, the vanishing regulator limit might indeed not feature any fixed point for the extended gravity-matter system. If this is the case, the inclusion of the higher-order, induced couplings is critical to see the lack of fixed point in this limit.\\
Second, our truncation might be too small, and additional higher-order couplings, which appear in the beta function for $g_{\phi}$ have to be taken into account to obtain the correct scaling of $G_{\rm crit.}$ with $a$.\\
Third, the vanishing regulator limit might be qualitatively different for dimensionless and dimensionful couplings: Given that in this limit the regulator vanishes, there is no requirement that fixed points must necessarily persist in this limit. It is a surprising sign of stability that the critical exponents for the dimensionless matter couplings stay finite and nonzero in this limit. Conversely, it is not necessarily a sign of a problem for the fixed point, that it does not persist for the dimensionful, induced couplings.

\section{Conclusions and outlook}\label{sec:conclusions}
In this paper, we demonstrate that one can obtain misleading results about matter-gravity systems by focusing on non-universal quantities. We do so by using the vanishing-regulator setup from \cite{Baldazzi:2020vxk,Baldazzi:2021ijd}. This provides us with a one-parameter family of results that depend on the parameter $a \in [0,1]$. For $a \rightarrow 1$, we recover standard FRG results. The limit $a \rightarrow 0$ is one in which the FRG regulator vanishes. Therefore, a priori, there is no reason to expect that a nontrivial flow remains. In this setup, we investigate the quantum-gravity contribution to the flow of a gauge coupling $g$, a Yukawa coupling $y$ and a scalar quartic coupling $\lambda$. We show that the gravitational contribution at a constant value of the Newton coupling goes to zero, when $a \rightarrow 0$. This appears to be in line with the expectation that the limit $a \rightarrow 0$ is trivial. It also fits to perturbative results on gravity-matter systems, which show a vanishing gravitational contribution to matter beta functions in the absence of a dimensionful regularization, e.g., when using dimensional regularization \cite{Toms:2007sk,Ebert:2007gf,Anber:2010uj}.\\
However, this changes when one evaluates the flow at the gravitational fixed point $G_{\ast}$, which itself is $a$-dependent, and diverges for $a \rightarrow 0$. This divergence is again in line with perturbative results on gravity, which require a dimensionful regularization for a finite gravitational fixed point to exist \cite{Niedermaier:2009zz}.\\
Combining the $a$ dependence of the gravitational fixed point with the $a$-dependence of the gravitational contribution to $\beta_g$, $\beta_y$ and $\beta_{\lambda}$, the $a \rightarrow 0$ limit becomes non-trivial and finite results are obtained in each case. This combined quantity corresponds to a critical exponent, i.e., a universal quantity. This is in contrast to the gravitational contribution at constant $G$, which is a non-universal quantity. This non-universal quantity has been in the focus of studies of beta functions in perturbation theory \cite{Robinson:2005fj,
Pietrykowski:2006xy,Toms:2007sk,Ebert:2007gf,Toms:2008dq,Tang:2008ah,Toms:2009vd,Toms:2010vy,Anber:2010uj,Ellis:2010rw,Toms:2011zza,Felipe:2011rs,Narain:2012te,Bevilaqua:2021uzk}. Our results serve to demonstrate that caution is required in the interpretation of these results. \\
Thus, our results call for a re-investigation of results in perturbation theory, in a setting where dimensionful regularizations can be removed in a controlled fashion, to discover, whether universal quantities (such as critical exponents at a fixed point) are nontrivial in perturbation theory, thus providing additional evidence for asymptotic safety in gravity-matter systems.

Our results apply both in the ''standard" gravity setting as well as in unimodular gravity, where the gravitational effect on matter has also been a previous focus of perturbative investigations \cite{Gonzalez-Martin:2017bvw}. In both settings, we find a nontrivial $a \rightarrow 0$ limit only if we focus on universal quantities, but not, if we evaluate the gravitational contribution to flows of matter couplings at constant gravitational coupling $G$.

Finally, our results also show that not all quantities in asymptotically safe gravity have a finite $a \rightarrow 0$ limit. We find that the critical value of the gravitational coupling which delineates the weak-gravity bound, diverges faster for $a \rightarrow 0$ than the gravitational fixed-point value. Thus, there are settings where a fixed point satisfies the weak-gravity bound at $a =1$, but violates it for $a \rightarrow 0$. This may be an artefact of our present truncation, or might indicate that the vanishing regulator limit is less suitable to study aspects of gravity-matter systems related to higher-order matter couplings.


\bigskip
\emph{Acknowledgements:} We thank Roberto Percacci and Alessio Baldazzi for useful discussions.
This work is supported by a grant (29405) by VILLUM fonden.

\appendix

\section{Steps beyond 1-loop approximation \label{app::Beyond1Loop}}

The analysis presented in Sec.~\ref{sec::Results_Universal} relies on the 1-loop approximation for the flow equation of the Newton coupling. As a result, the critical exponent associated with $G_*$ takes the value $\theta_G = 2$ for any value of $a$. This result is a trivial consequence of a flow equation of the form $k \pt_k G = 2 G + B(a) G^2$, irrespective of the details of $B(a)$. 

Here, we complement our analysis by adding contributions beyond the 1-loop approximation. We include contributions in $\beta_G$ that are proportional to the anomalous dimensions arising from the regulator insertion $\pt_t \textbf{R}_k$. The resulting flow equation for $G$ is given by
\begin{align}\label{eq::beta_G_improved}
	k \pt_k G = 2 G &+ 
	\frac{8\pi}{3} \big( 22 I_{1,1}[\hat{r}_a] + 11 \,I_{1,2}[\hat{r}_a]
	\big)  G^2 \nonumber \\
	&+ 
	\frac{8\pi}{3} \big( 12\,\eta_h I_{0,1}^1[\hat{r}_a]  
	-2 \eta_c I_{0,1}^1[\hat{r}_a] + 11 \eta_c I_{0,2}^1[\hat{r}_a] 
	\big) G^2,  
\end{align} 
with $\eta_h$ and $\eta_c$ denoting the graviton and ghost anomalous dimensions, respectively. For simplicity, we do not consider the impact of matter fields\footnote{For this reason the numerical factors in the $\eta$-independent terms of \eqref{eq::beta_G_improved} differ from the ones in \eqref{eq::Flow_G}.}. All the qualitative results remain unchanged in the presence of matter fields.

In the \textit{unimodular gravity} setting, we obtain the following flow equation for the Newton coupling 
\begin{align}\label{eq::beta_G_UG_improved}
	k \pt_k G |_\textmd{UG} = 2 G &+ 
	\frac{8\pi}{3} \big( 22 I_{1,1}[\hat{r}_a] - 6 \,I_{1,2}[\hat{r}_a]
	\big)  G^2 \nonumber \\
	&+ 
	\frac{4\pi}{3} \big( 24\,\eta_h I_{0,1}^1[\hat{r}_a]  - 15\,\eta_h I_{0,2}^1[\hat{r}_a] + 
	-4 \eta_c I_{0,1}^1[\hat{r}_a] + 18 \eta_c I_{0,2}^1[\hat{r}_a]
	\big) G^2 \,.
\end{align} 

We can compute $\eta_h$ and $\eta_c$ by applying a derivative expansion to the flow of graviton and ghost 2-point functions. We find the following results:
\begin{align}
	&\eta_h = \frac{ A_h \, G - ( A_h B_{c,2} - A_c B_{h,2}) G^2}
	{1 - (B_{h,1} + B_{c,2}) \,G + (B_{h,1} B_{c,2} - B_{h,2} B_{c,1}) \,G^2 }  \,, \label{eq::etah_NNLO} \\
	&\eta_c =  \frac{ A_c \, G - ( A_c B_{h,1} - A_h B_{c,1}) G^2}
	{1 - (B_{h,1} + B_{c,2}) \,G + (B_{h,1} B_{c,2} - B_{h,2} B_{c,1}) \,G^2 }  \,. \label{eq::etac_NNLO}
\end{align}
We provide an ancillary notebook containing the explicit form of the coefficients $A$ and $B$ both in the \textit{standard} and \textit{unimodular} settings.

We also investigate an intermediate approximation, where we expand $\eta_h$ and $\eta_c$ up to the first order in $G$, namely
\begin{align}
	&\eta_h = A_h \,G + \mathcal{O}(G^2) \,, \label{eq::etah_NLO}\\
	&\eta_c = B_h \,G + \mathcal{O}(G^2)   \, \label{eq::etac_NLO}.
\end{align}
This is equivalent to neglecting the anomalous dimensions arising from the regulator insertion $\pt_t \textbf{R}_k$ in the flow equations for the anomalous dimensions.

We use the following terminology for the different approximations we are considering: 
i) \textit{leading order} (LO) - this is obtained by setting $\eta_h=\eta_c=0$ in the flow equation for $G$ (Eqs. \eqref{eq::beta_G_improved} and \eqref{eq::beta_G_UG_improved}); 
ii) \textit{next to leading order} (NLO) - this is obtained by combining \eqref{eq::etah_NLO} and \eqref{eq::etac_NLO} with the flow equation for $G$; 
iii) \textit{next to next to leading order} (NNLO) - this is obtained by combining \eqref{eq::etah_NNLO} and \eqref{eq::etac_NNLO} with the flow equation for $G$.

\begin{figure}[!t]
	\begin{center}
		\hspace*{-.5cm}
		\includegraphics[height= 6.cm]{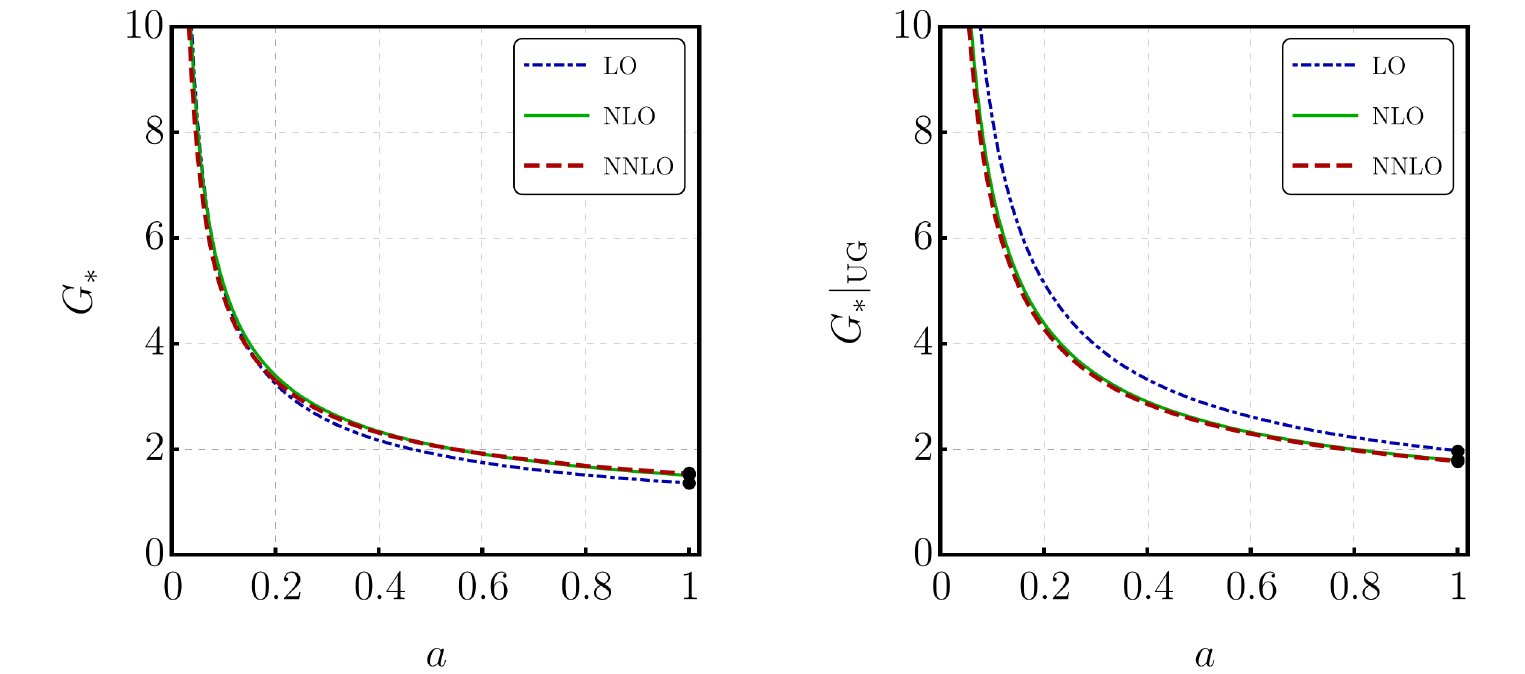} 
		\caption{We show the fixed point $G_*(a)$ both in standard gravity (left panel) and unimodular gravity (right panel). In both cases we consider different approximations for the anomalous dimensions $\eta_h$ and $\eta_c$.}
		\label{fig::G_fixpt_beyond1loop}
	\end{center}
\end{figure}

In Fig.~\eqref{fig::G_fixpt_beyond1loop}, we show the fixed-point value for $G$ in the three different approximations. 
We report our results both in \textit{standard} and \textit{unimodular} gravity, but we focus on results obtained with a Litim-type regulator. 
However, all the results reported in this section remain qualitatively the same for the exponential regulator.

As one can see in Fig.~\eqref{fig::G_fixpt_beyond1loop}, all the approximations that we considered lead to the same qualitative behavior. 
In particular, for small values of $a$, the fixed-point value for $G$ behaves as
\begin{align}\label{eq::Gfp_scaling_App}
	G_*(a) = - \frac{\mathcal{C}}{a \log (a)}  +  \Omega(a) \,,
\end{align}
with $\Omega(a)$ corresponding to finite contributions when $a \to 0$ and with $\mathcal{C}$ being a (positive) constant whose specific value depends on the choice of approximation.

The behavior observed in \eqref{eq::Gfp_scaling_App} remains the same if we include matter contributions to the flow of $G$ (at least for a small number of fermions and scalars). 
The scaling $G_*(a) \sim (a \log (a))^{-1}$ is precisely what is necessary to cancel out $a \log(a)$-contributions, producing finite results for $f_g$, $f_\lambda$ and $f_y$ in the vanishing regulator limit. 
Therefore, we can conclude that the results presented in Sec. \ref{sec::Results_Universal} allow an extension to NLO and NNLO approximations.

\begin{figure}[!t]
	\begin{center}
		\hspace*{-.5cm}
		\includegraphics[height= 6.cm]{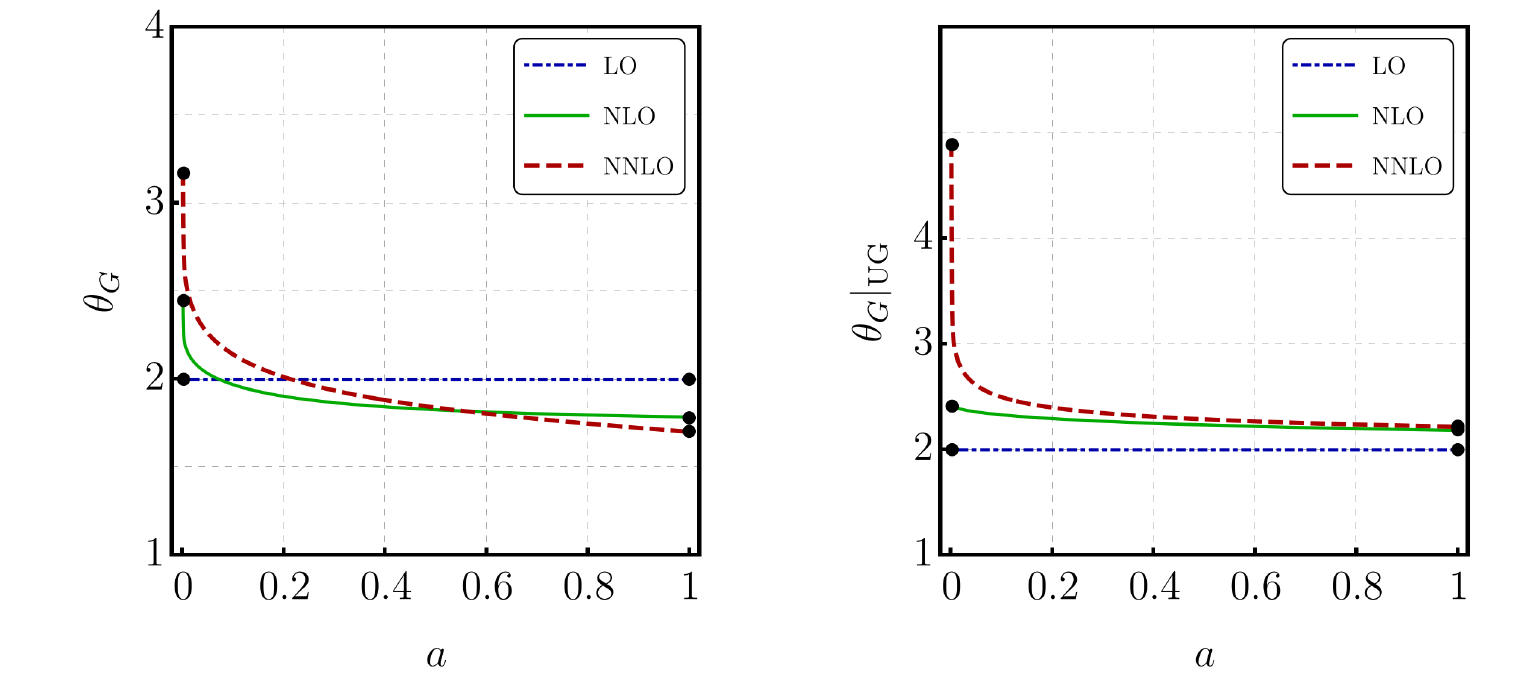} 
		\caption{We plot the critical exponents $\theta_G$ as a function of the interpolating parameter $a$. The left panel shows the results in standard gravity (left panel). The right panel shows the results in unimodular gravity. In both cases we consider different approximations for the anomalous dimensions $\eta_h$ and $\eta_c$.}
		\label{fig::Crit_Exp_beyond1loop}
	\end{center}
\end{figure}

In Fig.~\ref{fig::Crit_Exp_beyond1loop}, we show the behavior of the critical exponent associated with $G$ for the different approximations discussed here. 
As expected, the NLO and NNLO results deviate from the ``naive value'' $\theta_G = 0$. 
However, despite the divergent behavior of $G_*(a)$, we observe a finite value of $\theta_G$ in the vanishing regulator limit.  

\begin{figure}[!t]
	\begin{center}
		\hspace*{-.5cm}
		\includegraphics[height= 13.cm]{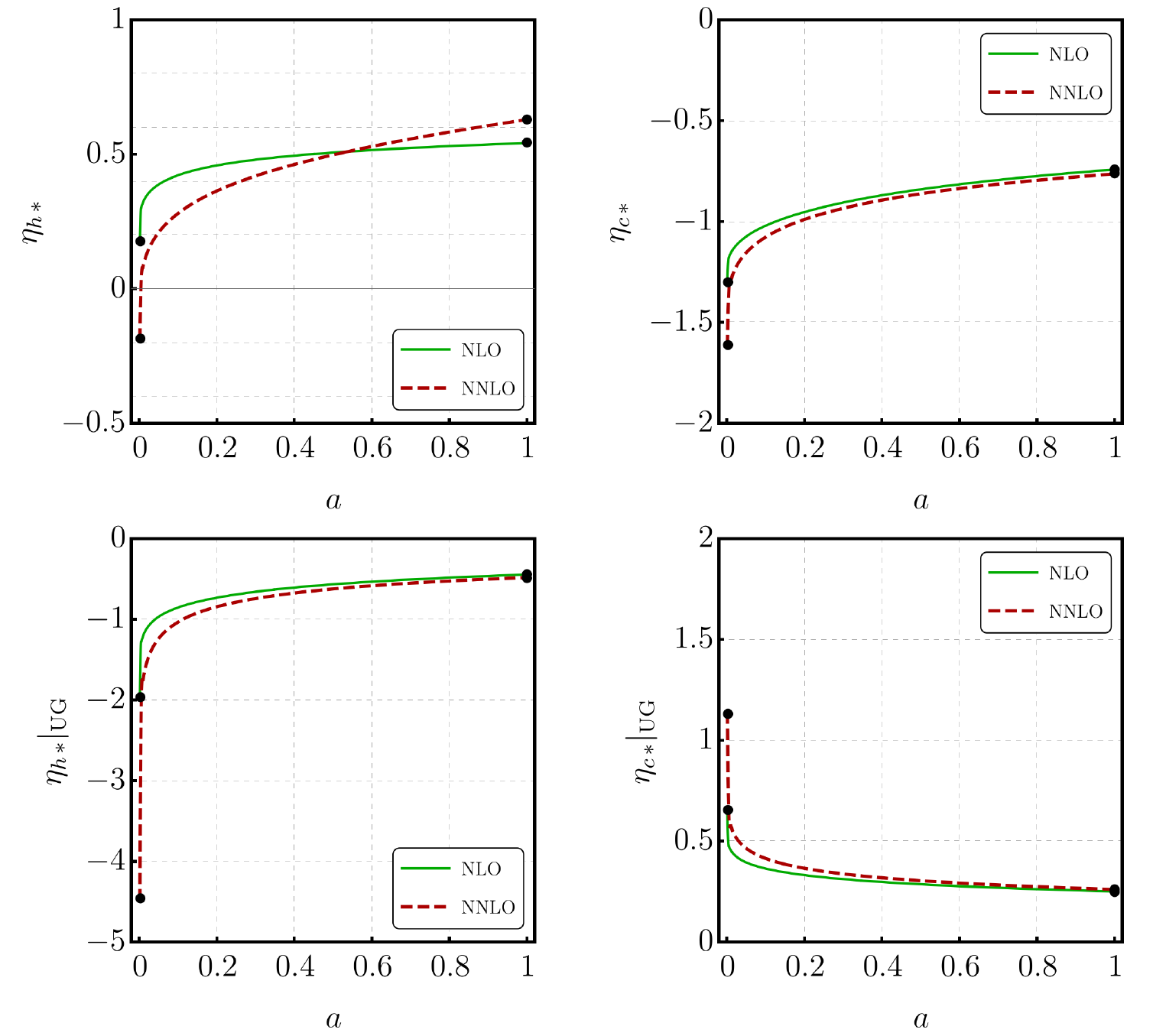} 
		\caption{We show the graviton and ghost anomalous dimensions evaluated at the fixed point $G_*$ as a function of the interpolating parameter $a$. In the first row we show the results for the \textit{standard gravity} setting, while in the second row we show the results for the \textit{unimodular gravity} setting.}
		\label{fig::Etas_beyond1loop}
	\end{center}
\end{figure}

Finally, in Fig.~\ref{fig::Etas_beyond1loop}, we show the graviton and ghost anomalous dimensions evaluated in the NLO and NNLO approximations. 
Once again, we observe non-trivial cancellations involving the scaling $G_*(a) \sim (a \log (a))^{-1}$, resulting in finite anomalous dimensions in the vanishing regulator limit.

\newpage
\bibliographystyle{JHEP}

\bibliography{references}
\end{document}